\documentclass[11pt,twoside,letter]{article}
\usepackage{amsthm, amsmath, amssymb, amsfonts}
\usepackage{mathtools}

\usepackage[normalem]{ulem}

\usepackage{bbm}
\usepackage{mathrsfs}
\usepackage{graphicx}

\usepackage[shortlabels]{enumitem}
\usepackage[colorlinks=true, pdfstartview=FitV, linkcolor=blue, citecolor=blue, urlcolor=blue]{hyperref}
\usepackage[usenames]{color}

\usepackage{url}
\makeatletter\def\url@leostyle{%
 \@ifundefined{selectfont}{\def\UrlFont{\sf}}{\def\UrlFont{\scriptsize\ttfamily}}} \makeatother

\usepackage[margin=1.0in, letterpaper]{geometry}

\newtheorem{theorem}{Theorem}

\theoremstyle{definition}

\theoremstyle{remark}
\newtheorem{remark}[theorem]{Remark}

\numberwithin{equation}{section}
\numberwithin{theorem}{section}

\definecolor{Red}{rgb}{0.9,0,0.0}
\definecolor{Blue}{rgb}{0,0.0,1.0}
\def\cF{\mathcal{F}}

\def\cT{\mathcal{T}}

\def\cW{\mathcal{W}}

\def\cZ{\mathcal{Z}}

\def\bE{\mathbb{E}}
\def\bF{\mathbb{F}}

\def\bR{\mathbb{R}}

\newcommand{\wh}{\widehat}

\newcommand{\1}{\mathbbm{1}}            
\newcommand{\set}[1]{\{#1\}}            
\renewcommand{\mid}{\;|\;}              
\newcommand{\norm}[1]{ \| #1 \| }       

\DeclareMathOperator{\Var}{\mathrm{Var}}          
\usepackage[table]{xcolor}
\usepackage{colortbl}

\title{Robo-Advising in Motion: A Model Predictive Control Approach}

\author{
	Tomasz R. Bielecki\,\thanks{Department of Applied Mathematics, Illinois Institute of Technology
		\newline \hspace*{1.45em}  10 W 32nd Str, Building RE, Room 220, Chicago, IL 60616, USA
		\newline \hspace*{1.45em}  Emails: \url{tbielecki@illinoistech.edu} (T. R. Bielecki), and \url{cialenco@illinoistech.edu} (I. Cialenco)
		\newline \hspace*{1.45em}  URLs: \url{http://math.iit.edu/\~bielecki}  and \url{http://cialenco.com}
		\vspace{0.5em}} ,
	\and
	Igor Cialenco\,\footnotemark[1]        
}

\date{ {\small 
		First circulated and this version: January 11, 2026\\
}}

\begin{document}

\maketitle

%
%
	\vspace{-2em}
	
	\smallskip
	
{\footnotesize
	
	\noindent\rule{\textwidth}{0.4pt}
	\begin{minipage}[t]{0.10\linewidth}
		\textsc{Abstract:}
	\end{minipage}%
	\hfill
	\begin{minipage}[t]{0.88\linewidth}
		Robo-advisors (RAs) are automated portfolio management systems that complement traditional financial advisors by
		offering lower fees and smaller initial investment requirements. While most existing RAs rely on static,
		one-period allocation methods, we propose a dynamic, multi-period asset-allocation framework that leverages Model
		Predictive Control (MPC) to generate suboptimal but practically effective strategies. Our approach combines a Hidden
		Markov Model with Black–Litterman (BL) methodology to forecast asset returns and covariances, and incorporates
		practically important constraints, including turnover limits, transaction costs, and target portfolio allocations.
		We study two predominant optimality criteria in wealth management: dynamic mean–variance (MV) and dynamic
		risk-budgeting (MRB). Numerical experiments demonstrate that MPC-based strategies consistently outperform myopic
		approaches, with MV providing flexible and diversified portfolios, while MRB delivers smoother allocations less
		sensitive to key parameters. These findings highlight the trade-offs between adaptability and stability in
		practical robo-advising design.
	\end{minipage}
	
	\vspace{1em}	\noindent
	\begin{minipage}[t]{0.10\linewidth}
		\textsc{Keywords:}
	\end{minipage}%
	\hfill
	\begin{minipage}[t]{0.88\linewidth}
		robo advising, model predictive control, dynamic asset allocation, mean-variance optimization,
		risk-budgeting, Black–Litterman model, Hidden Markov Model, wealth management, rolling horizon optimization
	\end{minipage}
	
		\vspace{0.7em}	\noindent
	\begin{minipage}[t]{0.10\linewidth}
		\textsc{MSC2020:}
	\end{minipage}%
	\hfill
	\begin{minipage}[t]{0.88\linewidth}
		Primary 91G10; Secondary 91G60
	\end{minipage}
	
	\vspace{0.7em}\noindent
	\begin{minipage}[t]{0.10\linewidth}
		\textsc{JEL:}
	\end{minipage}%
	\hfill
	\begin{minipage}[t]{0.88\linewidth}
		Primary G11; Secondary G17
	\end{minipage}
	\noindent\rule{\textwidth}{0.4pt}

}



\section{Introduction}  \label{sec:intro}

Robo-advisors (RAs) are fully or semi-automated portfolio management systems designed for individual investors, complementing traditional human financial advisors by offering lower fees and requiring smaller initial investments. These systems are becoming increasingly prevalent and influential in the wealth and asset management industry. For a comprehensive review and analysis of various aspects of robo-advisors, we refer to the extensive literature, including the works of Beketov et al. \cite{Beketov2018}, Rossi and Utkus \cite{RossiUtkus2020}, D’Acunto and Rossi \cite{DAcuntoRossi2021}, Helms et al. \cite{Helms2021}, Alsabah et al. \cite{AlsabahEtAl2020}, Capponi et al. \cite{CapponiEtAl2022} and the recent literature survey by Akhtar et al. \cite{Akhtaretal2025}. 

The key components of any robo-advising system are its portfolio selection methodology (asset allocation or management) and the interface through which the investor’s risk–return profile is elicited. To the best of our knowledge, all currently operating RAs rely on static, one-period portfolio allocation frameworks; see Beketov et al. \cite{Beketov2018} for a discussion of this point.
While the elicitation of an investor’s risk–return profile is extensively studied in the behavioral finance literature, its implementation in robo-advising is constrained by strict regulatory requirements, and in practice is typically carried out through standardized questionnaires; for more details cf.  \cite{CapponiEtAl2022} and reference therein.

In this work, we primarily focus on building an asset-allocation engine and consider several heterogeneous risk-return profiles representing standard investor types. We propose a dynamic (multi-period) framework for designing the asset-allocation engine of a robo-advisor. Motivated by insights from the broader literature on dynamic decision-making, we argue that a dynamic asset-management approach offers meaningful advantages over static (myopic) allocation methods in terms of overall portfolio performance.

Extending existing RA systems directly to a fully stochastic dynamic framework--whether in discrete or continuous time--is computationally infeasible. The main challenge lies in the fact that the underlying optimization criteria, together with practical constraints, lead to a time-inconsistent problem that cannot be efficiently solved over multiple periods. One approach to addressing such dynamic problems is to consider subgame-perfect (time-consistent) solutions (cf. \cite{BjoerkMurgoci2014}), or alternatively to construct a simplified approximate formulation that yields a suboptimal yet practically effective dynamic strategy.
Subgame-perfect solutions remain computationally demanding and, by their very nature, depend on the investment horizon; moreover, they must be reconstructed from scratch whenever the investor updates their risk attitude, making them less suitable for large-scale applications such as robo-advisors. Moreover, subgame-perfect solutions assume that the investor’s risk–return profile is known for all future times, which makes the resulting allocations harder to interpret and less practical for large-scale applications like robo-advisors.

Our proposed suboptimal approach is rooted in Model Predictive Control (MPC), a methodology widely used in engineering for dynamic decision problems under uncertainty (stochastic control problems). Rather than seeking a globally optimal solution, the essence of MPC is to approximate a stochastic control problem through a sequence of deterministic problems, solving only for the next few steps and updating as new information becomes available. This also means that the investor’s risk–return profile needs to be known only over a short-term horizon, with the allocation updated dynamically at each step. MPC is particularly suitable for robo-advisors primarily because the optimization problems it requires are numerically fast and tractable, often taking the form of convex programs.  A key ingredient of any MPC algorithm is the forecasting module. In our implementation, we combine a Hidden Markov Model (HMM) with elements of the Black-Litterman (BL) methodology to forecast mean returns and covariances for portfolio assets. The inclusion of a BL based component is inspired by Wealthfront, one of the earliest and most prominent robo-advisors, which uses BL estimates in its static asset-allocation engine. For background on HMM and BL methodologies, see, e.g., \cite{Zucchini2009} and \cite{Meucci2008}, respectively. Our numerical experiments show that this suboptimal MPC-based approach outperforms traditional myopic strategies.

We conduct our study of dynamic asset-allocation systems for RAs using two alternative optimality criteria: the dynamic mean–variance (MV) criterion and the dynamic risk-budgeting (MRB) criterion, which are among the predominant approaches in current wealth management. Unlike traditional utility-maximization frameworks, which are rarely applied in practical robo-advising, these criteria allow for straightforward and interpretable portfolio decisions. In our analysis, we also incorporate practically important constraints such as turnover limits, transaction costs, and target portfolio allocations to ensure realistic and implementable strategies.

The results indicate that the MPC-based suboptimal strategies consistently outperform myopic strategies, with particularly strong improvements under the MV criterion. In contrast to typical engineering applications, where MPC performance is highly sensitive to the key rolling-horizon parameter, we find that in our asset-allocation setup the results are much less sensitive, especially under the MRB criterion, where its inherent structure makes portfolio outcomes relatively stable across variations in this parameter.

The obtained results indicate that MPC-based suboptimal strategies consistently outperform myopic strategies, with particularly strong improvements under the MV criterion combined with BL framework by producing a more diversified and stable portfolios than classical estimated MV approaches. Nevertheless, MV with BL remains sensitive to noisy or rapidly changing risk profiles, making explicit turnover, transaction cost, and target portfolio constraints essential for stabilizing portfolio dynamics in practical robo-advising applications. In contrast, the MRB with strategy generates markedly smoother and more stable allocations across a wide range of parameters, including risk-aversion, transaction cost, and MPC horizon, reflecting its inherent robustness. While this stability reduces turnover, it also limits responsiveness to market forecasts, and soft transaction cost penalties are often ineffective, requiring hard constraints when trading activity must be controlled. Overall, these findings suggest that effective RA design should carefully align the choice of optimization criterion and structure of the imposed constraints.

The paper is organized as follows. In Section~\ref{sec:preliminaries}, we review relevant background material on MPC. Section~\ref{sec:SMPC-general} formulates SMPC problems tailored to the operations of RAs. Our approach for forecasting the conditional moments of asset returns is presented in Section~\ref{sec:forecast}. In Section~\ref{sec:risk-profiling}, we discuss the modeling of parameters related to the investor’s risk profile and trading constraints. Finally, in Section~\ref{sec:numerical}, we illustrate the proposed robo-advising methodology through a numerical example based on market data. Concluding remarks are detailed in Section~\ref{sec:conclusion}.

\section{Preliminaries on Model Predictive Control} \label{sec:preliminaries}

Model Predictive Control (MPC) is a methodology designed to provide sub-optimal, yet computationally efficient, solutions to a wide range of control problems. Numerical evidence widely demonstrates that the resulting loss of optimality is typically modest and represents an acceptable trade-off for the gain in computational efficiency. Extensive practical applications of MPC, particularly in engineering, demonstrated its strong empirical performance. To the best of our knowledge, there are currently no theoretical results that quantify the degree of sub-optimality of MPC.

Model Predictive Control (MPC) should be viewed as a general framework rather than a single, uniquely defined method. It admits a wide range of formulations and implementations, differing in modeling assumptions, objective functions, and constraints handling. We refer the reader to the literature for representative treatments and discussions; see, for example, \cite{Yan2021}, \cite{Hewing2020}, \cite{RawlingsEtAl2020}, \cite{RakovicLevine2019}, \cite{KouvaritakisCannon2016}, and references therein.


Here we present a tentative description of what can be considered Stochastic MPC (SMPC), which addresses stochastic control problems within the MPC framework.  We need to stress though that the terminology  ``stochastic model predictive control'' is not uniquely defined in the MPC universe.  Following the existing literature, we focus on the discrete-time setting.


On a given probability space $(\Omega,\cF,P)$, let us consider a prototypical discrete time controlled stochastic dynamical system:
\begin{align}\label{eq:dyn}
X_{\tau+1}=f(X_\tau,U_\tau,\epsilon_\tau),\ \tau=0,1,2,\ldots,
\end{align}
where $X$ is a $\bR^n$-valued state process, $U$ is a $\bR^m$-valued control process, and $\epsilon$ is a $\bR^k$-valued driving, underlying random process. As is customary, we assume that additional constraints are imposed on the controls
\begin{align}\label{eq:cons}
U_\tau\in \mathcal{U}_\tau,\ \tau=0,1,2,\ldots ,
\end{align}
where $\mathcal{U}_\tau=\mathbf{U}_\tau(X_\tau)\subset \bR^m$ for some measurable multifunction $\mathbf{U}_\tau$.

Fix a time horizon $T>0$ and define a (running) optimization criterion
\[
	J_t(X,U,T) = \sum_{\tau=t}^{T} F(X_\tau, U_\tau), \quad t=0,\ldots, T,
\]
where $F$ is an appropriate real-valued function. This criterion forms the basis of the objective that we aim to optimize within the MPC.

Let $\mathbb{F}=(\mathcal{F}_\tau)_{\tau=0,1,2,...}$ denote some relevant filtration modeling the flow of information available to the controller, which is non-anticipative w.r.t. process $\epsilon$.  Typically, $\mathcal{F}_\tau=\sigma(X_s,\, s=0,1,\ldots,\tau)$, but not necessarily - see Remark \ref{rem:3-2}. We write $Y_\tau \in \mathcal{F}_t$ to indicate that the random variable $Y_\tau$ is $\mathcal{F}_t$-measurable. Throughout, $E$ will denote the expectation under $P$, and we put $E_\tau[\cdot]:= E[\cdot\mid \cF_\tau]$. Respectively, $\Var$ and $\Var_\tau$ will denote the variance and conditional variance.

The classical exemplary stochastic control problem takes the form
\begin{align}\label{eq:SCP1}
V_0:=\sup_{U_\tau\in \mathcal{G}_\tau,\, \tau=0,1,2,...,T-1 }E[J_0(X,U,T)|\mathcal{F}_0],
\end{align}
subject to \eqref{eq:dyn} and \eqref{eq:cons}.

Throughout, we assume that the necessary conditions are satisfied by the model primitives, such as measurability, and that all processes are adapted to $\bF$.  We further assume the existence of an optimal control sequence $U^{*}$ for the above problem, so that
\[
V_0 = {E}\big[J_0(X,U^{*},T) \mid \mathcal{F}_0\big].
\]
We refer to \cite{BertsekasShreve1978Book} for a general treatment of stochastic control problems in discrete time.

A well-founded theoretical way to solve the above problem is to use dynamic programming (DP), for problems that satisfy the dynamic programming principle, also called time consistent problems. As it is well documented though, practical application of dynamic programming is typically prohibited by the so called "curse of dimensionality," which leads to formidable computational difficulties. In case of problems that do not satisfy the dynamic programming principle, the dynamic programming will not work as a method for delivering the optimal solution.

A theoretically well-founded approach to solve the above problem is dynamic programming (DP), applicable to problems that satisfy the dynamic programming principle, also referred to as time-consistent problems. However, as widely documented, practical application of DP is often limited by the so-called ``curse of dimensionality,'' which results in formidable computational challenges. For time-inconsistent problems, the DP cannot be used in principle to obtain the optimal solution, and usually some other methods are employed such as sub-game perfect solutions, or adaptive change of the criteria.

MPC was introduced as a methodology to generate sub-optimal solutions to the above stochastic control problem that can be computed efficiently, either in cases where dynamic programming is inapplicable or where its application is hindered by the curse of dimensionality.

Next, we introduce the SMPC formulation for the problem \eqref{eq:dyn}–\eqref{eq:SCP1}, suitable for our robo-advising applications, and consider a receding-horizon approach, where the initial time and horizon advance sequentially. Specifically, one chooses a rolling horizon $H>0$,\footnote{Typically, $H\leq T$.} and for $t=0,1,\ldots, T-1,$ one solves
\begin{align}\label{eq:F-hat}
\sup_{U_\tau\in \mathcal{G}_t,\, \tau=t,t+1,...,t+H-1 }\sum_{\tau=t}^{t+H-1}\widehat F_t(U_t,\ldots,U_\tau),
\end{align}
subject to \eqref{eq:dyn} and \eqref{eq:cons}, where $\widehat F_t(U_t,\ldots,U_\tau)$ is the prediction (or forecast) of $F(X_\tau,U_\tau)$ based on the information carried by $\mathcal{F}_t$.  Thus, the problem \eqref{eq:F-hat} is essentially a deterministic and open-loop control problem in discrete time as all the controls $U_\tau$ are constrained to be measurable with respect to the initial information $\mathcal{F}_t$.  Once the optimal controls, say $U^*_{t,\tau},\, \tau=t,t+1,...,t+H-1,$ are computed at time $t$, only the control  $U^*_{t,t}$ is implemented at time $t$. Then, the process repeats at time $t+1$, and so on. The model predictive controls generated in this way are
\[
U^{\textrm{MPC}}_\tau:=U^*_{\tau,\tau},\ \tau=0,1,\ldots,T-1.
\]
With additional structure imposed on $\wh F_t$, such as convexity, the SMPC solution reduces to a sequence of deterministic optimization problems that are relatively easy to solve and can, in practice, be computed on the fly.

\section{SMPC for Robo-Advisors}\label{sec:SMPC-general}

An RA, as an automated system, aims to manage a portfolio of financial assets either fully autonomously or semi-autonomously on behalf of the investor, with the objective of optimizing the investor’s risk–reward profile. This process typically involves periodic interactions between the RA (machine) and the investor, primarily to calibrate the risk–reward criterion in a way that best aligns with the investor’s preferences.

We propose to cast the RA’s automated decision-making within the SMPC framework.
There are three key components of such framework:
\begin{itemize}  \addtolength{\itemsep}{-2pt}
\item Risk-reward criterion, which in turn involves selection of a) investment horizon; b) investor's risk profile; c) RA's attitude towards transaction costs or turn-over. We consider two main criteria: the mean-variance criterion, and the mean-risk-budgeting criterion.
\item Forecasts related to the asset returns. We will use a combination of Hidden Markov Model and Black-Litterman methodologies for this purpose.
\item Constraints imposed on the composition and re-balancing of the asset portfolio. The main constraints we consider are the long-only positions, self-financing and turn-over limits.
\end{itemize}

We assume that the RA invests in a portfolio consisting of $N$ assets.  We denote by $\pi_0$ the vector of holding proportions (portfolio weights) at time $0$ (before re-balancing at time $0$), and for $\tau=0,1,\dots$ we denote by  $\pi_{\tau+1}$ the vector of holding proportions  at time $\tau$  \textbf{after re-balancing}. Thus, for $\tau=0,1,\dots$,
\[\sum_{i=1}^N\pi_\tau^i=1.\]
Additionally, for $\tau=0,1,\dots$, we consider two trading constraints that are often used in practice:
\begin{itemize}
\item The \textit{long only constraint}, that is $\pi_\tau\geq 0$.
\item The \textit{turnover constraint} $\|\pi_{\tau+1}-\pi_{\tau}\|_1=\sum_{i=1}^N|\pi^i_{\tau+1}-\pi^i_{\tau}|\leq \delta$ for some positive constant $\delta$.
\end{itemize}

We denote by $r_\tau=(r^1_\tau,\ldots,r^N_\tau)$ the (a priori) \textbf{random} vector of returns on the $N$ assets between times $\tau$ and $\tau+1$, $\tau=0,1,\ldots\, $. That is,
\[r^i_\tau=\frac{P^i_{\tau+1}-P^i_{\tau}}{P^i_{\tau}},\]
where $P^i_{\tau}$ and $P^i_{\tau+1}$ are the prices of asset $i$ at time $\tau$ and $\tau+1$, respectively.

We denote by $\mathbb{F}=(\mathcal{F}_\tau,\ \tau=0,1,\ldots)$ the filtration given as: $\mathcal{F}_0=\{\emptyset,\Omega\}$ and $\mathcal{F}_\tau=\sigma(r_0,\ldots,r_{\tau-1})$ for $\tau=1,2,\ldots $.

Let $W_\tau$ denote the wealth of the portfolio as time $\tau=0,1,\dots$. Under the self-financing constraint we have
\[W_\tau=\sum_{i=1}^N {\pi}^i_\tau P^i_{\tau}= \sum_{i=1}^N {\pi}^i_{\tau+1}P^i_{\tau}.\]

In addition, if the trading is self-financing then the following relationship holds between return on the portfolio and the returns on the portfolio constituents
\[\frac{W_{\tau+1}-W_\tau}{W_\tau} =\sum_{i=1}^N r^i_\tau \pi^i _{\tau+1} =r_\tau^\intercal \pi_{\tau+1}.\]

\subsection{Mean-variance portfolio selection problem}\label{sec:MV-selection}

One of the cornerstone risk-reward criteria used in financial markets is the classical mean-variance (MV) problem pioneered by Harry Markowitz, and  later modified by Fisher Black and Robert Litterman. It is well-known that the stochastic mean-variance problems are time inconsistent, and do not satisfy the dynamic principles (cf. \cite{BjoerkMurgoci2014, LiNg2000, BCC2020} for a comprehensive treatment). Moreover, for mean-variance problems with short-selling constraints does not have an explicit solution even in one-period time setup. Nevertheless, this criterion remains widely used by wealth managers, typically in a myopic setting, where optimization is performed at each time step using a one-period objective. In this work, we propose an RA that adopts a dynamic mean–variance criterion integrated with the Black–Litterman methodology within a SMPC framework.

For motivational purposes, we first consider a \textit{one-period investment problem} defined over the interval from the initial time time $\tau=0$ and the terminal time $T=\tau+1=1$. Let $\pi_0$ be the $\mathcal{F}_0$ measurable vector of portfolio weights at time $0$ (before rebalancing at time $0$), so that $\1^\intercal\pi_0=1$, and let $\pi_1$ be a $\mathcal{F}_0$  measurable vector of portfolio weights at time $1$ (after rebalancing at time $0$), so that $\1^\intercal\pi_1=1$. Without loss of generality we assume that $\cF_0$ is trivial. The (random) return on the portfolio between times $0$ and $1$ is $r_0^\intercal\pi_1$. Note that  $\pi_0$  and $\pi_1$ are deterministic here, and $\pi_1$ is chosen at time $\tau=0$, without knowledge of $r_0$.

The the classical MV criteria amounts to
\[
E_0[r_0^\intercal\pi_1] -\gamma_0 \Var_0[r_0^\intercal\pi_1],
\]
for some constant $\gamma_0>0$, that quantifies the risk attitude of the investor at time $\tau=0$.

To account for trading costs, the first level modification of this criterion is
\[
E_0[r_0^\intercal\pi_1] -\gamma_0 \Var_0[r_0^\intercal\pi_1]-\eta_0 TC\left(\pi_1-\pi_0\right ),
\]
where $TC$ is a function representing trading costs,\footnote{The trading costs are comprehensively described in terms of a trading cost function, say $TC$. The form of this function may vary between specific applications.} and $\eta_0 >0$ is the \textit{trading-costs sensitivity factor}.

Since $E_0[r_0^\intercal\pi_1]=E_0[r_0]^\intercal\pi_1$, and $\Var_0[r_0^\intercal\pi_1]=\pi^\intercal_1\Sigma_0\pi_1,$, with $\Sigma_0=\Var_0[r_0]$, the corresponding one-period MV asset management problem is
\begin{equation}\label{MV1-0}
\sup_{\pi_1\in \mathcal{F}_0}\Big ( E_0[r_0]^\intercal\pi_1 -\gamma_0 \pi^\intercal_1\Sigma_0\pi_1-\eta_0 TC\left(\pi_1-\pi_0\right )\Big),
\end{equation}
subject to constraints on $\pi_1$: $\pi_1\geq 0$, $\1^\intercal \pi_1=1$, and  $\|\pi_1-\pi_0\|_1\leq \delta$.

\bigskip

The generalization of the above single-period problem to the \textit{multi-period case} is straightforward.  For $\tau=0,1,\ldots$, let $\cF_\tau $ be the information used by an investor at time $\tau$ to rebalance their portfolios from $\pi_\tau$ to $\pi_{\tau+1}$ at time $\tau$. Thus $\pi_{\tau},\, \tau=0,1,\ldots, $ is a random process adapted to filtration $\mathbb{F}$.

An important consideration needs to be given to the portfolio rebalancing frequency. In this paper we assume that portfolio can be rebalanced on a fixed set of deterministic dates. Denote the set of rebalancing dates staring at time $\tau=0$ as $\mathbf{R}(0)$. For example, if $\tau$ is a daily time scale, and  $\mathbf{R}(0)=\set{0,7,14,21,\ldots}$ then rebalancing takes place every seven days. By $\mathbf{NR}(0)$ we denote the set of non-rebalancing dates starting at time $\tau=0$. In the example above $\mathbf{NR}(0)=\set{1,2,3,4,5,6,8,9,10,\ldots,13,15,\ldots}$.

The multi-period  MV problem (referring to the portfolio returns) is
\begin{align}\label{SPE1}
\sup_{\pi_{\tau+1}\in \mathcal{F}_\tau,\, \tau =0,\ldots,T-1}\sum_{\tau=0}^{T-1}\Big (E_0[r_{\tau}^\intercal\pi_{\tau+1}] -\gamma_0 \Var_0[r_{\tau}^\intercal\pi_{\tau+1}]- \eta_0 E_0[TC\left(\pi_{\tau+1}-\pi_{\tau}\right )]\Big),
\end{align}
where $\pi_0$ is given and satisfies $\1^\intercal \pi_0=1,\ \pi_0\geq 0,$ and subject to constraints on $\pi_1,\ldots,\pi_T$,
\begin{align}\label{SPE3}
 & \pi_{\tau+1} \geq 0, \ \1^\intercal \pi_{\tau+1} =1,\ \|\pi_{\tau+1}-\pi_\tau\|_1\leq \delta,\ \tau=0,...,T-1, \\
& \pi_{\tau+1}=\pi_\tau, \ \tau\in \mathbf{NR}(0). \label{SPE4}
\end{align}
Additional trading constraints, such as turn-over budget or target portfolio and we refer to \cite{Boyd2017} for a comprehensive discussion; specifically, see Section 4.4 (the set $\cW_t$ of holding constraints), Section 4.5 (the set  $\cZ_t$ of trading constraints), and Section 4.6 (soft constraints). These include a wide range of constraints and costs, many of which are also discussed in, for example, Roncalli \cite{Roncalli2016Book}. In this work, we deliberately restrict attention to a limited but fundamental set of trading constraints, corresponding to those most commonly implemented in practical robo-advisor systems. In addition to turn-over constraint, in Section~\ref{sec:target-portfolio} we also consider target portfolio constraint.

\begin{remark}\label{rem:3-2}
Note that the problem  \eqref{SPE1}-\eqref{SPE4} can be written as \eqref{eq:SCP1}, by taking
$\epsilon_\tau=r_{\tau+1}$, $U_{\tau}=\pi_{\tau+1}$, $X_0=r_0$,
$f(X_{\tau},U_\tau,\epsilon_\tau)=\epsilon_\tau$,
$\mathcal{G}_\tau =\mathcal{F}_\tau,$
$$
F_\tau(X_\tau,U_\tau)=E_0[X_{\tau}^\intercal U_\tau]-\gamma_0 \Var_0[X_{\tau}^\intercal U_\tau)]-\eta_0 TC\left(U_{\tau}-U_{\tau-1}\right ),
$$
and
\[
{\mathbf U}_\tau(X_\tau)=\{
 U_{\tau} \geq 0,\ \1^\intercal U_{\tau} =1,\ \|U_{\tau}-U_{\tau-1}\|_1\leq \delta,\ U_{\tau}=U_{\tau-1},\ \tau\in \mathbf{NR}(0)\},
\]
for $\tau=0,1,\ldots T-1, $   where $U_{-1}=\pi_0$.
\end{remark}

\subsubsection{ MV receding horizon SMPC}\label{sec:MV-SMPC}

Assume that the RA has a methodology to predict or forest the returns and variance at each time instant, and we  denote by $\widehat r_{\tau|t}$ the forecast of the mean of $r_{\tau}$, and by $\widehat \Sigma_{\tau|t} $ the forecast of the variance-covariance matrix of $r_{\tau}$ based on the information available to the RA at time $t\leq \tau$. As mentioned earlier, the used forecasting methodologies is a key component in our framework, and we study it in Section~\ref{sec:forecast}.

We fix an integer valued \textit{SMPC rolling horizon} $H>0$. The receding horizon SMPC corresponding to problem \eqref{SPE1}-\eqref{SPE4} is then stated as follows:

\smallskip \noindent
\textbf{Step A.} {Set $t=\min \mathbf{R}(0)$ to be the smallest element of the set of rebalancing dates $\mathbf{R}(0)$.} Go to Step B.

\smallskip \noindent
\textbf{Step B.} Choose $\gamma_t$ and $\eta_t$\footnote{We refer to Section~\ref{sec:risk-profiling} for a discussion of the choice of $\gamma_t$ and $\eta_t$.}, and solve the following problem
\begin{align}\label{SMPC1-new-trb}
\sup_{\pi_{\tau+1}\in \mathcal{F}_t,\, \tau=t:t+H-1}\sum_{\tau=t}^{t+H-1}\Big (\widehat r_{\tau|t}^\intercal\pi_{\tau+1}  -\gamma_t \pi_{\tau+1}^\intercal \widehat \Sigma _{\tau|t}\pi_{\tau+1}-\eta_t TC\left(\pi_{\tau+1}-\pi_\tau\right )\Big),
\end{align}
 where $\pi_t$ is given and satisfies $\1^\intercal \pi_t=1,\ \pi_t\geq 0$, and subject to constraints on $\pi_{t+1},\ldots,\pi_{t+H}$,
\begin{align}\label{SMPC3-trb}
& \pi_{\tau+1} \geq 0,\ \1^\intercal \pi_{\tau+1} =1,\ \|\pi_{\tau+1}-\pi_{\tau}\|_1\leq \delta,\ \tau=t,...,t+H-1, \\
\label{SMPC4-trb}
&\pi_{\tau+1}=\pi_{\tau}, \ \tau\in \mathbf{NR}(t),
\end{align}
where $\mathbf{NR}(t)$ is the set of no-rebalancing dates consistent with $\mathbf{NR}(0)$ in the sense that $\mathbf{NR}(t)=\mathbf{NR}(0)\cap\mathbf{NR}(t)$.

Denote by $\pi^*_{\tau+1|t},\ \tau=t,...,t+H-1,$ the optimal controls for problem \eqref{SMPC1-new-trb}--\eqref{SMPC3-trb}. Define the model predictive control for time $t$ as
\[\pi^{MPC}_{t+1}:=\pi^*_{\tau+1|t},\]
and apply this control to the system at time $t$. That is, chose your rebalanced portfolio at time $t$ according to $\pi^{MPC}_{t+1}$. Go to Step C.

\smallskip \noindent
\textbf{Step C.} Set {$t:=\min \mathbf{R}(t+1)$}. If $t\leq T-1$ then go to Step B. If $t>T-1$ then stop.

\bigskip
If the rebalancing is done monthly, say every 30 days, then $\min \mathbf{R}(t+1)=t+30$. Of course, in applications a more adequate rebalancing tenor must be applied, consistent with the calendar of trading days.

We remark that similar to one-period problem, since it is required that $\pi_{t+1},\ldots,\pi_{t+H}$ are all $\cF_\mathbf{t}$ measurable, the above SMPC procedure is a sequence of open-loop control problems. Moreover, if one takes the function $TC$ to be convex, then the control problems in Step B are (static) convex optimization problems are static convex optimization problems for which highly efficient off-the-shelf solvers are available.

Note that in the above SMPC the counterpart of $\widehat F_t(U_t,\ldots,U_\tau)$ showing in \eqref{eq:F-hat} is
\[
\widehat r_{\tau|t}^\intercal\pi_{\tau+1}  -\gamma_t \pi_{\tau+1}^\intercal \widehat \Sigma _{\tau|t}\pi_{\tau+1}-\eta_t TC\left(\pi_{\tau+1}-\pi_\tau\right ).
\]

\subsection{Mean-risk-budgeting (MRB) portfolio selection}\label{sec:example-risk-budgeting}
Another widely studied approach to portfolio construction is risk budgeting, including its special case of risk parity, in which portfolio weights are determined so that each asset or asset class contributes a predetermined proportion of total portfolio risk; cf. \cite{Roncalli2016Book}. We adapt this approach to the SMPC setting.

Let $\rho$ be a risk measure assessing the risk of portfolio $\pi_{\tau+1}$. Specifically, we take the standard deviation based risk measure
\[
\rho(\pi_{\tau+1})=\sqrt{\Var_0[r_\tau^\intercal\pi_{\tau+1}]}.
\]
It is well known that this risk measure satisfies the Euler risk allocation principle, under which the contribution of the $i$-th asset to the total aggregated risk is given by
\[
{\mathcal{RC}}^i_\tau=\pi_{\tau+1}^i\partial_i\rho(\pi_{\tau+1}).
\]
In the risk budgeting paradigm, the risk contributions are required to match prescribed proportions; thus, for each asset $i=1,\ldots,N$, we aim to achieve
\[
{\mathcal{RC}}^i_\tau=b_i\rho(\pi_{\tau+1})=b_i\sqrt{\Var_0[r_\tau^\intercal\pi_{\tau+1}]},
\]
or, equivalently,
\[
{\mathcal{MRB}}^i_\tau:=\frac{\pi_{\tau+1}^i\partial_i\rho(\pi_{\tau+1})}
{\sqrt{\Var_0[r_\tau^\intercal\pi_{\tau+1}]}}=b_i,
\]
where $b_i>0$ are given and such that $\sum_{i=1}^N b_i=1$.

\begin{remark}\label{rem:gammaR} (i) By analogy to the mean-variance criteria, we propose to parameterize the risk-budgeting weights $b_i$ by a risk attitude parameter $\gamma_R$ as follows.
Divide the assets in two classes by their riskiness, e.g. bonds and equity. Denote by $N_B$ the number of assets in the first class, and assign equal risk-budget weights $b_i= \gamma_R/(N_B(1+\gamma_{R}))$ to all assets in the first class, and correspondingly, $b_i=1/((N-N_B)(1+\gamma_{R}))$ for the assets in the second class. Thus if the second class will contain riskier assets, then a larger $\gamma_R$ will budget smaller weight on the risky assets, and thus indeed the coefficient $\gamma_R$  plays the role of the attitude-aversion or risk-tolerance coefficient  of the investor. We use this approach in defining the risk-budgeting weights $b_i$ in Section~\ref{sec:numerical}, and show how $\gamma_R$ changes the structure of the portfolio and its characteristics.

By similarity, assets may be partitioned into several classes, with weights assigned at the class level and equal weights within each class.

\smallskip \noindent
(ii)
A special case of the risk-budgeting method, the so called \textit{risk parity paradigm}, is to assign equal risk weights to all assets, that is for each $i=1,\ldots,N,$
\[
\frac{\pi_{\tau+1}^i\partial_i\rho(\pi_{\tau+1})}
{\sqrt{\Var_0[r_\tau^\intercal\pi_{\tau+1}]}}=\frac{1}{N}.
\]
In particular, note that $\gamma_{R}=(N-N_B)/N_B$ corresponds to the risk-parity case.
\qed
\end{remark}

The multi-period  MRB problem  (referring to the portfolio returns) is

\begin{align}\label{SPE1-MRB}
\sup_{\pi_{\tau+1}\in \mathcal{F}_\tau,\, \tau =0,\ldots,T-1}\sum_{\tau=0}^{T-1}\Big (E_0[r_\tau^\intercal\pi_{\tau+1}]
 -{\phi_\tau} \sum_{i=1}^{N}({\mathcal{MRB}}^i_\tau-b_i)^2- \eta_\tau E_0[TC\left(\pi_{\tau+1}-\pi_\tau\right )]\Big),
\end{align}
where $\pi_0$ is given and satisfies
$ \1^\intercal \pi_0=1,\ \pi_0\geq 0,$ and subject constraints on $\pi_1,\ldots,\pi_T$,
\begin{align}\label{SPE3-MRB}
& \pi_{\tau+1} \geq 0,\ \1^\intercal \pi_{\tau+1} =1,\ \|\pi_{\tau+1}-\pi_\tau\|_1\leq \delta,\ \tau=0,...,T-1,\\
\label{SPE4-MRB}
&\pi_{\tau+1}=\pi_\tau,\ \tau\in \mathbf{NR}(0).
\end{align}

We emphasis that while the mean–variance optimization balances expected return against overall portfolio risk (measured as variance), the risk-budgeting allocates portfolio weights so that each asset contributes proportionally to total risk. The coefficient $\phi_\tau$, called the \textit{risk-budgeting sensitivity coefficient}, is not directly tied to the investor's risk profile; rather, it quantifies the degree to which the risk-budgeting allocation is enforced.

\subsubsection{MRB receding horizon SMPC}\label{sec:SMPC-MRB}

Let $\widehat \rho(\pi_{\tau+1}|t)$ be a forecasted value of the risk measure assessing the risk of portfolio $\pi_{\tau+1}$, based on information in $\mathcal{F}_t$. Specifically, postulating that $\pi_{\tau+1}\in \mathcal{F}_t$, we take  $\widehat \rho(\pi_{\tau+1}|t)=\sqrt{\pi^\intercal_{\tau+1} \widehat \Sigma_{\tau|t}\pi_{\tau+1}} $. Consequently, the forecasted Euler contribution of the $i$-th asset to the total risk is given as
\[
\widehat {\mathcal{RC}}^i_{\tau|t}=\frac{(\pi_{\tau+1})^i (\widehat \Sigma _{\tau|t}\pi_{\tau+1})^i}{\sqrt{\pi^\intercal_{\tau+1} \widehat \Sigma _{\tau|t}\pi_{\tau+1}}}.
\]
Accordingly, the forecasted risk-budgeting term is
\begin{equation}\label{eq:ith-raw} \widehat {\mathcal{MRB}}^i_{\tau|t}=\frac{(\pi_{\tau+1})^i (\widehat \Sigma _{\tau|t}\pi_{\tau+1})^i}{\pi^\intercal_{\tau+1} \widehat \Sigma _{\tau|t}\pi_{\tau+1}}=\frac{\pi_{\tau+1}^\intercal \widehat \Sigma^{(i)}_{\tau|t}\pi_{\tau+1}}{\pi^\intercal_{\tau+1} \widehat \Sigma _{\tau|t}\pi_{\tau+1}},
\end{equation}
where as before the super index $i$ indicates the $i$-th coordinate of a vector, and where $\Sigma^{(i)}_{\tau|t}$ is the matrix obtained from $\Sigma_ {\tau|t}$ by leaving the $i$-th row intact and replacing all other elements in $\Sigma_{\tau|t}$ by zeros.

The time $t$ optimization problem becomes
\begin{align}\label{SMPC1-new-trb-MRB}
\sup_{\pi_\tau\in \mathcal{F}_t,\, \tau=t+1,\ldots,t+H}\sum_{\tau=t}^{t+H-1}\Big (\widehat r_{\tau|t}^\intercal(\pi_{\tau+1}) -\phi_t \sum_{i=1}^{N}( \widehat {\mathcal{MRB}}^i_{\tau|t}-b_i)^2-\eta_t TC\left(\pi_{\tau+1}-\pi_{\tau}\right )\Big),
\end{align}
subject to the same constraints \eqref{SMPC3-trb}-\eqref{SMPC4-trb}.

The MBR receding horizon SMPC procedure follows the same steps as in MV case detailed in Section~\ref{sec:MV-SMPC}, with \eqref{SMPC1-new-trb} replaced by \eqref{SMPC1-new-trb-MRB}.

\subsubsection{Convex approximation to the MRB criterion}\label{sec:SMPC-MRB-Convex}

The criterion in \eqref{SMPC1-new-trb-MRB} is generally non-convex. To enable efficient numerical solution, following \cite{LI2021}, we introduce a convex approximation of this criterion, which retains the essential features of \eqref{SMPC1-new-trb-MRB} while allowing for tractable optimization.

Towards this end, we  define
\begin{align*}
d_{\tau|t,i}(\pi_{\tau+1})&:=\widehat {\mathcal{MRB}}^{i}_{\tau|t}-b_i,\\
d_{\tau|t}(\pi_{\tau+1})&:=[d_{\tau|t,1}(\pi_{\tau+1}),\ldots,d_{\tau|t,N}(\pi_{\tau+1})]^\intercal,
\end{align*}
and we proceed according to the following recursion:\\
\underline{Iteration $k=0$.} Initialize an arbitrary allocation $\pi^0=(\pi^0_\tau,\, \tau=t+1,\ldots,t+H)$, and fix $\rho^0=[0,1]$ and a tolerance $tol>0$.

\smallskip \noindent
\underline{Iteration $k$.}
For any iteration $k=0,1,\ldots $ we expand $d_{\tau|t,i}(\pi_{\tau+1})$ around $\pi^k$:
\[
d_{\tau|t,i}(\pi_{\tau+1})\approx d_{\tau|t,i}(\pi^k_{\tau+1})+\left (\nabla d_{\tau|t,i}(\pi^k_{\tau+1})\right )^T\left (\pi_{\tau+1}-\pi^k_{\tau+1}\right ),
\]
where $\nabla d_{\tau|t,i}$ is the gradient. The convex approximation to the criterion  \eqref{SMPC1-new-trb-MRB} becomes
\begin{align}
\sup_{\pi_\tau\in \mathcal{F}_t,\, \tau=t+1,\ldots,t+H}&\sum_{\tau=t}^{t+H-1}\Bigg [\widehat r_{\tau|t}^\intercal\pi_{\tau+1} \nonumber
-\phi_t \Bigg \{\sum_{i=1}^{N}\Bigg ( d_{\tau|t,i}(\pi^k_{\tau+1}) 
+\left (\nabla d_{\tau|t,i}(\pi^k_{\tau+1})\right )^T\left (\pi_{\tau+1}-\pi^k_{\tau+1}\right )\Bigg )^2 \\
&\qquad \qquad +\frac{\kappa_\tau}{2}\|\pi_{\tau+1}-\pi^k_{\tau+1}\|^2_2\Bigg \}-\eta_t TC\left(\pi_{\tau+1}-\pi_{\tau}\right )\Bigg ], \label{SMPC1-new-trb-MRB-convex}
\end{align}
where $\frac{\kappa_\tau}{2}\|\pi_{\tau+1}-\pi^k_{\tau+1}\|^2_2$ is a regularization term with some $\kappa_\tau>0$, and where we take
$TC$ to be convex, e.g. $TC\left(\pi_{\tau+1}-\pi_{\tau}\right )= \|\pi_{\tau+1}-\pi_{\tau}\|_1$ and see Section~\ref{sec:tr_costs}.

We will now proceed to rewrite the criterion in \eqref{SMPC1-new-trb-MRB-convex} in a more convenient  way, for which we note that
\begin{align*}	
&\sum_{i=1}^{N}\left ( d_{\tau|t,i}(\pi^k_{\tau+1})+\left (\nabla d_{\tau|t,i}(\pi^k_{\tau+1})\right )^T\left (\pi_{\tau+1}-\pi^k_{\tau+1}\right )\right )^2
+\frac{\kappa_\tau}{2}\|\pi_{\tau+1}-\pi^k_{\tau+1}\|^2_2 \\
& \qquad \qquad =\frac{1}{2}\pi^\intercal(\tau+1)Q^k_{\tau+1|t}\pi_{\tau+1} +\pi^\intercal(\tau+1)q^k_{\tau+1|t}+ \textrm{constant},
\end{align*}
where (cf. \cite{LI2021}, Section 4.2.1)
\begin{align*}
Q^k_{\tau+1|t}&=2(A^k_{\tau+1|t})^\intercal A^k_{\tau+1|t}+\kappa_\tau I, \\
q^k_{\tau+1|t}&=2(A^k_{\tau+1|t})^\intercal d_{\tau|t}(\pi^k_{\tau+1})-Q^k_{\tau+1|t}\pi^k_{\tau+1}, \\
A^k_{\tau+1|t}&=[\nabla d_{\tau|t,1}(\pi^k_{\tau+1}),\ldots,\nabla d_{\tau|t,N}(\pi^k_{\tau+1})]^\intercal
\end{align*}
\begin{align*}
\nabla d_{\tau|t,i}(\pi^k_{\tau+1})=\frac{(\pi^k_{\tau+1})^\intercal \widehat \Sigma _{\tau|t}\pi^k_{\tau+1}\left(\widehat \Sigma^{(i)}_{\tau|t}+\left (\widehat\Sigma^{(i)}_{\tau|t}\right )^\intercal\right )\pi^k_{\tau+1}}{\left((\pi^k_{\tau+1})^\intercal \widehat \Sigma _{\tau|t}\pi^k_{\tau+1}\right )^2} \\
 -2\frac{(\pi^k_{\tau+1})^\intercal \widehat \Sigma^{(i)} _{\tau|t}\pi^k_{\tau+1}\widehat \Sigma _{\tau|t}\pi^k_{\tau+1}}{\left((\pi^k_{\tau+1})^\intercal \widehat \Sigma _{\tau|t}\pi^k_{\tau+1}\right )^2},
\end{align*}
and where $\widehat \Sigma^{(i)} _{\tau|t}$ is defined right below \eqref{eq:ith-raw}

Now, restate  the problem  \eqref{SMPC1-new-trb-MRB} as
\begin{align}\label{SMPC1-new-t-MRB-quad}
\sup_{\pi_\tau\in \mathcal{F}_t,\, \tau=t+1,\ldots,t+H}&\sum_{\tau=t}^{t+H-1}-\phi_t\Big (\frac{1}{2}\pi^\intercal_{\tau+1}Q^k_{\tau+1|t}\pi_{\tau+1} +\pi^\intercal_{\tau+1}q^k_{\tau+1|t}\Big)\\&-\eta_t \|\pi_{\tau+1}-\pi_\tau\|_1+\pi^\intercal_{\tau+1}\widehat r_{\tau|t}, \nonumber
\end{align}

subject to constraints on $\pi_{t+1},\ldots,\pi_{t+H}$
\begin{align}\label{SMPC3-MRB-quad}
& \pi_{\tau+1} \geq 0,\ \1^\intercal \pi_{\tau+1} =1,\ \|\pi_{\tau+1}-\pi_{\tau}\|_1\leq \delta,\ \tau=t,...,t+H-1, \\
\label{SMPC4-trb-quad}
&\pi_{\tau+1}=\pi_\tau, \ \tau\in \mathbf{NR}(t).
\end{align}

Denote by $\pi^{*,k}_{\tau|t},\ \tau=t+1,...,t+H,$ the optimal controls for problem \eqref{SMPC1-new-t-MRB-quad}-- \eqref{SMPC4-trb-quad}.
Check stopping criterion: if
\[
\textrm{the improvement in the objective of Problem \eqref{SMPC1-new-trb-MRB}, \eqref{SMPC3-trb}-\eqref{SMPC4-trb}}  \leq \ tol,
\]
then
declare the model predictive control for time $t$ as
\[\pi^{MPC-MRB}_{t+1|t}:=\pi^{*,k}_{t+1|t},\]
and apply this control at time $t$.

Otherwise,
\begin{itemize}\addtolength{\itemsep}{-5pt}
\item Set $\pi^{k+1}_\tau=\pi^k_\tau+\rho^k(\pi^{*,k}_{\tau|t}-\pi^k_\tau),\ \tau=t+1,...,t+H,$
\item update $\rho^k$ to $\rho^{k+1}$: $\rho^{k+1}=\rho^{k}(1-\xi \rho^{k})$ where $\xi\in [0,1]$,
\item set $k\leftarrow k+1$, and go to Iteration $k$.
\end{itemize}

\section{Forecasting of conditional moments of asset returns via Hidden Markov Models and the Black-Litterman Methods}\label{sec:forecast}

This section addresses the computation of the forecasts $\widehat r_{\tau|t}$ and $\widehat \Sigma_{\tau|t}$, emphasizing methods that are both computationally efficient and practically accurate. Our approach is based on ideas from the theory of Hidden Markov Models (HMM) and from the Black-Litterman (BL) methodology frequently used in financial asset management.
There are a plethora of references regarding both HMM and Black-Litterman  methodologies, and we refer to \cite{NystrupEtAl2017,NystrupEtAl2018}  for the discussion of HMM relevant to our work, and to \cite{Meucci2008} for a discussion of the Black-Litterman model.

We deviate from the BL methodology by dispensing with the reverse optimization part that in the original BL model computes the expectation vector of the prior distribution of the portfolio means by backing it up from the market weights (see, e.g., \cite{Meucci2008}  formula (5)). Instead, we take as the expectation vectors of the prior distribution of the portfolio means, in each regime, the means vectors generated by the HHM module of our methodology.

As before, we assume that the  coefficients $\gamma_t$, $\phi_t$ and $\eta_t$ are given. The forecasting procedure proceeds according to the following stages.

\bigskip\noindent
\textbf{HMM forecasts of moments of asset returns.}
Our HMM forecasts are done using daily data. 
The asset management is done on the $\mathbf{R}(0)$ basis. In particular, portfolio rebalancing is done on the dates from the set $\mathbf{R}(0)$.

We assume that the vectors of returns $r_t,\ t=0,1,2,...,$ follow multivariate normal distributions.  As in \cite{LI2021} we consider two regimes of the economy: \textit{normal regime} and \textit{contraction regime}, with the respective prior distributions of returns being  $\mathcal{N}(\mu_n,\Sigma_n)$ and $\mathcal{N}(\mu_c,\Sigma_c)$. We take a HMM $(R_t,r_t)$, $t=0,1,2,...,$ where $R_t$ is the unobserved economy  regime at time $t$, and $r_t$ is the observed vector of returns at time $t$. Correspondingly, we assume that $R$ is a time-homogeneous Markov chain with the state space $\{n,c\}$, and with transition probability matrix
\[
\Lambda =
\begin{bmatrix}
    p_{nn} & 1-p_{nn} \\
    1-p_{cc} & p_{cc} \\
  \end{bmatrix} .
\]

The forecasts related to asset returns will be needed at each date $t$.  Accordingly, the HMM parameters $p_{nn}$, $p_{cc}$,  $\mu_n$, $\mu_c$, $\Sigma_n$ and $\Sigma_c$ can and will be updated at any time $t$, as new observations come between time $t-1$ and time $t$. We denote these updates as $p_{nn}(t)$, $p_{cc}(t)$,  $\mu_n(t)$, $\mu_c(t)$, $\Sigma_n(t)$ and $\Sigma_c(t)$. Then, the first two moments of the asset returns are estimated for future periods $\tau = t+1, \ldots, t+H$, by considering that model parameters are constant in $\tau$ (for a given $t$). For instance,  \cite{NystrupEtAl2018} considered time-$d$ dependent HMM parameter estimated using an online version of the expectation-maximization algorithm introduced by Stenger et al. (2001). Whereas in \cite{LI2021}, HMM parameters are fitted at every time $d$ based on returns of past 2000 days.

Now, given the information available at time $t$, i.e. on day $t$, we can update (or train) the HMM  to obtain the probability $q_t$ that the market is in the normal regime at time $t$. Based on that, and based on the time-$d$ fitted HMM parameters,  we compute the forecasts of the moments of prior distributions of returns. These forecasts are computed based on the postulate that the distribution of $r(\tau)$, given the information available at time $t$, is a mixture of distributions $\mathcal{N}(\mu_n,\Sigma_n)$ and $\mathcal{N}(\mu_c,\Sigma_c)$ with weights $q_{\tau|t}$ and $1- q_{\tau|t}$, where $ q_{\tau|t}$ is the probability  that the market is in the normal regime on day $\tau$ given the information available on day $t$. Thus, we have,
\begin{align}
q_{\tau|t}&=q_{\tau-1|t}p_{nn}(t)+(1- q_{\tau-1|t})(1-p_{cc}(t)),\ \tau=t+1,\ldots, \   q_{t|t}=q_t, \ \tau=t,t+1,\ldots, .\nonumber
\end{align}
In addition, for $\tau=t,t+1,\ldots , $ the first two forecasted moments of the mixture distribution are
\begin{align}
 \mu_{\tau|t}&= q_{\tau|t}\mu_{n}(t)+(1- q_{\tau|t})\mu_{c}(t), \\
 \Sigma_{\tau|t}&= q_{\tau|t}\Sigma_n+(1- q_{\tau|t})\Sigma_c+ q_{\tau|t}(\mu_n(t)\mu^\intercal_n(t)- \mu_{\tau|t} \mu^\intercal_{\tau|t}) \nonumber \\ &+(1- q_{\tau|t})(\mu_c(t)\mu^\intercal_c(t)- \mu_{\tau|t} \mu^\intercal_{\tau|t}).\nonumber
\end{align}

Note that the distribution of $r_\tau$, given the information available at time $t$, is not a Gaussian distribution, as it is a mixture of two Gaussians. We denote by $ r_{\tau|t}$ a random variable with the mixed Gaussians distribution as above.

\bigskip \noindent
\textbf{BL-like prior means of asset returns.} We now take the Bayesian view and we postulate that the means in each leg of the distribution of $r_{\tau|t}$, given the information available at time $t$, are random and distributed according to normal distributions. Specifically, to model the prior means
we let $\textbf{w}(t)$ be the vector of equilibrium weights at time $t$, and we let $\bar \lambda_{n/c}(t)$ be the average normal market risk-aversion as seen at time $t$ in regime $n/c$. Then, we postulate  that for
$\tau = t, \ldots, t+H-1,$ the prior means of the asset returns are\footnote{See e.g. \cite{Meucci2008} equations (2) and (5) for motivation.}
\begin{equation}\label{BL-prior-weights}
\mu^{BL,c}_{\tau|t}=2\bar \lambda_c(t) \Sigma_{\tau|t}\textbf{w}(t)+ \varepsilon^{\mu,c}(\tau| t),\ \mu^{BL,n}_{\tau|t}=2\bar \lambda_n(t) \Sigma_{\tau|t}\textbf{w}(t)+ \varepsilon^{\mu,n}_{\tau|t},
\end{equation}
where $\varepsilon^{\mu,c/n}_{\tau|t}$, $\tau = t, \ldots, t+H-1$ are independent noises, with $\varepsilon^{\mu,c/d}_{\tau|t} \sim \mathcal{N}(0, \iota_{c/n}{\Sigma} _{\tau  | t})$, and fixed parameters $\iota_{c/n}>0.$

We remark that in Black and Litterman paper \cite{BlackLitterman1992} the counterpart of $\iota_{c/n}$ is denoted by $\tau$ and indicates the uncertainty of the CAPM prior and ranges between 0.01 and 0.05. Other authors recommend choosing $\tau\in (0,1).$ See the discussion in \cite{Meucci2008} regarding the value assigned to $\tau$. The extreme value  $0$ of $\iota_{c/n}$ reflects degenerate prior: no uncertainty about the market implied prior.

\bigskip\noindent
\textbf{BL views.} In the spirit of the Black-Litterman methodology we now introduce the RA's views regarding the future values of expected returns on the $N$ assets included in our portfolios. The views are represented by deterministic $1\times K$ vectors denoted as
\begin{align*}
 v^{BL,c}_{\tau | t}  \ \textrm{and}\  v^{BL,n}_{\tau | t}.
\end{align*}
To simplify the set-up we have assumed that the number $K$ of views does not depend either on $t$ or $\tau$.  Again, in the spirit of the Black-Litterman methodology that relationship between the views and the BL prior means of asset returns is given as
\begin{align*}
 P_{c}\mu^{BL,c}_{\tau | t}&=v^{BL,c}_{\tau | t} + \varepsilon^{v,c}_{\tau | t}, \\ 
  P_{n}\mu^{BL,n}_{\tau | t}&=v^{BL,n}_{\tau | t} + \varepsilon^{v,n}_{\tau | t}, \\
  \tau &= t, \ldots, t+H-1,
\end{align*}
where $P_{c}$ and $P_{n}$ are $K\times N$ pick matrices, and where $\varepsilon^{v,c/n}_{\tau | t}$, $\tau = t, \ldots, t+H-1$ are independent noises, representing the degree of confidence granted to the view, with $\varepsilon^{v}_{\tau | t} \sim \mathcal{N}(0, \alpha_{c/n} {\Sigma}_{\tau | t})$. To simplify the set-up we have assumed that the pick matrices $P_{c}$ and $P_{n}$ do not depend either on $t$ or on $\tau$.

We note that $\alpha_{c/n} {\Sigma}_{\tau | t}$ corresponds to $\Omega$, and $\alpha_{c/n} $ corresponds to $\frac{1}{c}$ in \cite{Meucci2008}, where $c\in (0,\infty)$. Two extreme values of $\alpha_{c/n} $ are $0$ and $\infty$. The value $0$ reflects total confidence in the views, whereas $\infty$ reflects total lack of confidence in the views.

There is vast literature regarding generating views for the purpose of the classical Black-Litterman model. See e.g. \cite{Wutsqa2016,Geyer2016,Kara2019,Rezaei2021,BaruaSharma2022} and the references therein.
We will illustrate our approach with very special choices for the views and the pick matrices. Specifically, we consider the case of total absolute views where $K=N$, $P_c=P_n=I_{N\times N}$, with views chosen as
\begin{equation}\label{BL-view-HMM}
v_{BL,c}(\tau | t)=v_{BL,n}(\tau | t)= \mu(\tau|t).
\end{equation}

Other choices for the views and the pick matrices will be studied in a follow-up work.

\bigskip \noindent
\textbf{Posterior forecasts of the moments through Bayesian updating.} \textit{In the first step}, we do the Bayesian updating on each leg $c$ and $n$ for time $\tau$.  Let us take the $c$ leg.\footnote{Analogous formulae and remarks will apply to the $n$ leg.} Given the prior as in \eqref{BL-prior-weights} and the value of the views is $v^{BL,c}_{\tau| t}$  the posterior for this leg is Gaussian with mean
\begin{align*} 
 \widehat r^{BL,c}_{\tau | t}   &  = 2\bar \lambda_c(t) \Sigma_{\tau|t}\textbf{w}(t) \\
 & \qquad + \frac{\iota_{BL,c}}{\iota_{BL,c}+\alpha_{c}}{\Sigma}_{\tau  | t}P^\intercal_{c}\Bigg( P_{c}{\Sigma} _{\tau | t}P^\intercal_{c}\Bigg )^{-1}\Bigg (v^{BL,c}_{\tau| t}-P_{c}2\bar \lambda_c(t) \Sigma_{\tau|t}\textbf{w}(t)\Bigg )
\end{align*}
and variance-covariance matrix
\begin{align*}
 \widehat \Sigma^{BL,c}_{\tau | t} = (1+ \iota_{BL,c}){\Sigma}_{\tau  | t}-\frac{\iota^2_{BL,c}}{ \iota_{BL,c}+\alpha_{c} }{\Sigma} _{\tau  | t}P^\intercal_{c}\Bigg(P_{c}{\Sigma} _{\tau | t}P^\intercal_{c}\Bigg )^{-1}P_{c}{\Sigma} _{\tau  | t}).
\end{align*}

\begin{remark} Consider the prior as in \eqref{BL-prior-weights}. Then:
	\begin{enumerate}[(i)]\addtolength{\itemsep}{-2pt}
		\item In the case of no uncertainty of the prior, that is  $ \iota_{BL,c}=0$, we have that  $\widehat r^{BL,c}_{\tau | t}    = 2\bar \lambda_c(t) \Sigma_{\tau|t}\textbf{w}(t) $, and $\widehat \Sigma^{BL_c}_{\tau | t} = {\Sigma}_{\tau  | t}$.
		\item   In the case of total lack of confidence in the views, that is $\alpha_{c}=\infty$ we have $\widehat r^{BL,c}_{\tau | t}    = 2\bar \lambda_c(t) \Sigma_{\tau|t}\textbf{w}(t) $ , and $\widehat \Sigma^{BL_c}_{\tau | t} = (1+ \iota_{BL,c}){\Sigma} _{\tau  | t}$. The presence of the factor $(1+ \iota_{BL,c})$ reflects the estimation bias (error) in the prior.
		\item In the case of complete confidence in the absolute views, that is $\alpha_{c}=0$ and $P_{c}=Id$,  we have $\widehat r^{BL,c}_{\tau | t}    =v^{BL,c}_{\tau| t}$ and $\widehat \Sigma^{BL_c}_{\tau | t} = {\Sigma} _{\tau  | t}$.
	\end{enumerate}

\end{remark}

\textit{In the next step}, we mix the two posteriors to get the posterior forecasts of the moments of $r(\tau)$ as
\begin{align*}
\widehat r^{BL}_{\tau | t}&=q_{\tau|t}\widehat r^{BL,n}_{\tau | t} +(1-q_{\tau|t})r^{BL,c}_{\tau | t} , \\
\widehat \Sigma^{BL}_{\tau | t}&=q_{\tau|t}\widehat \Sigma^{BL,n}_{\tau | t}+(1-q_{\tau|t})\widehat \Sigma^{BL,c}_{\tau | t}  \\ &+q_{\tau|t}(\widehat r^{BL,n}_{\tau | t}(\widehat r^{BL,n}_{\tau | t})^\intercal-\widehat r^{BL}_{\tau | t}(\widehat r^{BL}_{\tau | t})^\intercal) \\
&+(1-q_{\tau|t})(\widehat r^{BL,c}_{\tau | t}(\widehat r^{BL,c}_{\tau | t})^\intercal-\widehat r^{BL}_{\tau | t}(\widehat r^{BL}_{\tau | t})^\intercal). 
\end{align*}

\begin{remark}\label{rem:SMPC-HMM-BL}
With these forecasting methodologies at hand, we conclude with the refined formulation of the MV and MRB approaches to RA using SMPC with HMM-BL. For the MV criterion (Section \ref{sec:MV-SMPC}) and, respectively, for the MRB criterion (Section~\ref{sec:SMPC-MRB}), we replace $\widehat r_{\tau|t}$ and $\widehat \Sigma_{\tau|t}$ with the BL forecasts $\widehat r^{BL}{\tau|t}$ and $\widehat \Sigma^{BL}{\tau|t}$, respectively.

\end{remark}

\section{Risk profiling,  allocation targets and trading constraints}\label{sec:g-t}

In this section we discuss how to model parameters related the investor's risk profile and the trading constraints.

\subsection{Risk profiling}\label{sec:risk-profiling}

One of the key features of a robo-advisor system is the possibility for the agent to interact with the robot on regular basis and reveal or update to the robot her current risk-reward preferences  that the robot has to use for future portfolio re-balances. There is a growing literature on how to elicit the risk preferences of an agent, and we refer to e.g. \cite{Dai2021,AlsabahEtAl2020,CapponiEtAl2022,CuiEtAl2022} and references therein. However, traditionally, but also following the regulatory frameworks requirements\footnote{In US under the SEC guidelines (SEC, 2006; SEC, 2019), and in Europe under  MiFID (Markets in Financial Instruments Directive) regulation.} the advisors are required to risk profile the  investors through a questionnaire. The outcome of such questionnaire is mapped to a numerical score, that we will identify in our model as the risk aversion coefficient denoted by $\gamma_t$, for $t\in\cT$. We assume that $\widetilde\gamma_t$ can take a finite number of values $\Gamma=\set{\gamma^1,\widetilde \gamma^2,\ldots, \gamma^N}$,  where we assume that  $ \gamma^1<  \gamma^2<\cdots<\gamma^N$.

For the MV criterion, risk preferences are encoded by the coefficient $\gamma_t \in \Gamma$, which directly parametrizes the risk--return trade-off in the objective.  For the MRB criterion, investor risk preferences are specified through the enforcement parameter $\phi_t$ and the risk-budgeting weights $b_i$, with $b_i$ parametrized by the coefficient $\gamma_R$ as described in Remark~\ref{rem:gammaR}.
The parameter $\phi_t$ governs the penalization strength of deviations from the prescribed risk budgets and is typically assumed constant over time.
The effective risk mapping in the MRB formulation is induced by $\gamma_R \in \Gamma_R$, which determines the allocation of aggregate risk across asset classes. The admissible set $\Gamma_R$ is defined as a mapping of $\Gamma$, ensuring consistency between the MV and MRB risk parametrizations.

In the numerical experiments, we consider several specifications for the risk-profile processes $\gamma_t$ and $\gamma_{R,t}$, which in a robo-advisory context represent the time evolution of the client’s risk profile as inferred by the platform:
\begin{itemize}
	\item \emph{Static profile}: $\gamma_t = \gamma_{R,t} = \gamma$ for all $t \in \mathcal{T}$. This specification corresponds to a client with a fixed risk profile and is used for most of the strategy comparison results;
	\item \emph{Lifecycle profile}: $\gamma_t \leq \gamma_{t+1}$ and $\gamma_{R,t} \leq \gamma_{R,t+1}$, capturing age-based de-risking commonly implemented in robo-advisors, whereby inferred risk tolerance decreases monotonically over time. In this study we use a linear decreasing function;
	\item \emph{Noisy profile}: $\gamma_t \sim \mathrm{Uniform}(\Gamma)$ and $\gamma_{R,t} \sim \mathrm{Uniform}(\Gamma')$, mimicking an inconsistent or unstable client profile arising from noisy questionnaires, behavioral biases, or frequent reassessment.
\end{itemize}
Other modeling approaches for the risk-aversion coefficient, including stochastic or learning-based methods, could also be considered but are beyond the scope of this study.

\subsection{Allocation target constraints}\label{sec:target-portfolio}
Another popular approach among financial advisors is to design a strategy that targets a specified portfolio at maturity, matching the investor's goal or target, with a specific maturity time. Usually, a more conservative portfolio the longer the horizon. Instead of targeting individual assets, we adopt a class-level approach: only the aggregate weights of asset classes, defined by risk profile, are constrained.
For instance, we partition the assets into two groups: $G_B$ consisting of fixed-income or lower-risk assets, and $G_S$ consisting of higher-risk assets such as equities.
We then specify target weights for each class, $w_B$ and $w_S$.
The target portfolio must satisfy the constraints\footnote{If there are only two groups, a single group constraint suffices since portfolio weights sum to one.}:
\begin{equation}\label{eq:target_group}
	\sum_{i\in G_B} \pi_\tau^i = w_B, \quad \sum_{i\in G_S} \pi_\tau^i = w_S.
\end{equation}

The class weights $w_B$ and $w_S$ are computed analogously to the MRB group weights (Remark~\ref{rem:gammaR}), i.e.,
\[
w_B = \frac{\gamma_{\text{target}}}{1 + \gamma_{\text{target}}}, \quad w_S = \frac{1}{1 + \gamma_{\text{target}}}.
\]

To avoid abrupt adjustments, these constraints are enforced gradually rather than only at the final step $\tau = T-1$.
This is achieved by linearly interpolating the risk-profile coefficient from the initial $\gamma$ to the target $\gamma_{\text{target}}$ and imposing \eqref{eq:target_group} for $\tau = T', \ldots, T$, where $T'$ is chosen appropriately.

\subsection{Transaction costs}\label{sec:tr_costs}
Transaction costs are a critical consideration in portfolio management, especially for algorithmic or robo-advising strategies.
They can be controlled through two complementary mechanisms.

First, through the \textit{soft penalty term} $\mathrm{TC}$ that we included in the objective function to discourage excessive trading. In this study, we take $TC(z)=||z||_1$, with various values for the scale parameters $\eta$.
A more comprehensive transaction cost function $\textrm{TC}$ is a version of the formula given in \cite{Boyd2017},
$TC(z)=\sum_{i=1}^N TC_i(z_i),$ where
$TC_i(z_i)=a_i|z_i|+ c_iz_i$, that allows to account for asymmetry in trading. We did not find significant qualitative changes in our numerical experiment  when using more complicated $\textrm{TC}$ functions, and we do not report these results here.

Second, \textit{hard constraints} on portfolio turnover can is imposed to explicitly limit the magnitude of changes in positions between consecutive periods, $\sum_{i=1}^{N} |\pi_{\tau+1}^i - \pi_\tau^i| \leq \delta$, $\tau = 0, \ldots, T-1$, where $\delta>0$ is the turnover bound. This combined approach allows the strategy to balance portfolio adjustments with the cost of trading, which is essential for realistic implementation in a robo-advisory framework.

A less common approach among robo-advisors, though occasionally used by portfolio managers, is a \emph{turnover-budget constraint}, which limits the total amount of transaction costs over the investment horizon.
We omit this feature in the present study, as we consider the two mechanisms discussed above--soft penalty term $\mathrm{TC}$ and hard turnover constraints--sufficient for practical transaction-cost management.
Nevertheless, given the path-dependent nature of turnover-budget constraints, their theoretical analysis could provide valuable insights.

\subsection{Initial portfolio}

 There is no canonical method to specify the initial portfolio, and in the literature common choices include equal weights, random weights and then ignoring several initial trades.
 For concreteness, we construct the initial portfolio based on an initial risk-profile coefficient, denoted $\gamma_{\text{IW}}$, by setting
 $\pi_0^i = b_i$, where $b_i$ are defined as in the MRB approach (Remark~\ref{rem:gammaR}).
 In most experiments, we take $\gamma_{\text{IW}} = \gamma_0$.

\section{Numerical experiment on market data}\label{sec:numerical}

In this section, we examine the proposed robo-advising methodology through a numerical example using market data. For our portfolio, we include eight Exchange-Traded Funds (ETFs), as detailed in Table~\ref{tab:etfs_by_category}. Common to mainstream robo-advisors, we have chosen a diversified portfolio of ETFs that spans various financial instruments (stocks, bonds, real estate) and different markets (developed, emerging). The data sets consist of daily closing asset prices (source: Yahoo! Finance) and daily total assets under management (source: ycharts.com) from January 2009 to December 2024, total 3268 data in the time series. The data were cleaned for outliers, and the time series were aligned to match dates with missing values accounted for.   Historical asset prices are shown in Figure~\ref{fig:all-prices}. The key statistics for each included asset and S\&P500 market index are displayed in Table~\ref{tab:key-stats}, and Table~\ref{fig:corr-matrix}

For these experiments, we fix January 2, 2020 as the starting trading day when the hypothetical client enters the market and first interacts with the RA. The RA re-balances portfolios monthly (every 22 days), for 56 periods of time; see gray shaded area in Figure~\ref{fig:all-prices}. Historical data from the past five years is utilized for calibration and estimation, applying a rolling window approach.

\begin{table}[h!]
	\centering
	\begin{tabular}{ll l}
		\hline
		\rowcolor{lightgray!30}
		Stocks & VTI &Vanguard Total Stock Market ETF \\
		\rowcolor{lightgray!30}	& VEA &Vanguard FTSE Developed Markets ETF \\
		\rowcolor{lightgray!30}	& VWO &Vanguard FTSE Emerging Markets ETF \\[0.3em]
		\rowcolor{lightgray!60}		Bonds & BND& Vanguard Total Bond Market ETF \\
		\rowcolor{lightgray!60}	& EMB& iShares J.P. Morgan USD Emerging Markets Bond ETF \\
		\rowcolor{lightgray!60}	& MUB &iShares National Muni Bond ETF \\
		\rowcolor{lightgray!60}	& LQD &iShares iBoxx \$ Investment Grade Corporate Bond ETF \\[0.3em]
		\rowcolor{lightgray!30}	Alternatives & VNQ& Vanguard Real Estate ETF \\
		\hline
	\end{tabular}
	\caption{List of included ETFs, by category. }
	\label{tab:etfs_by_category}
\end{table}

\begin{figure}[h]
	\centering
	\includegraphics[width=0.8\textwidth]{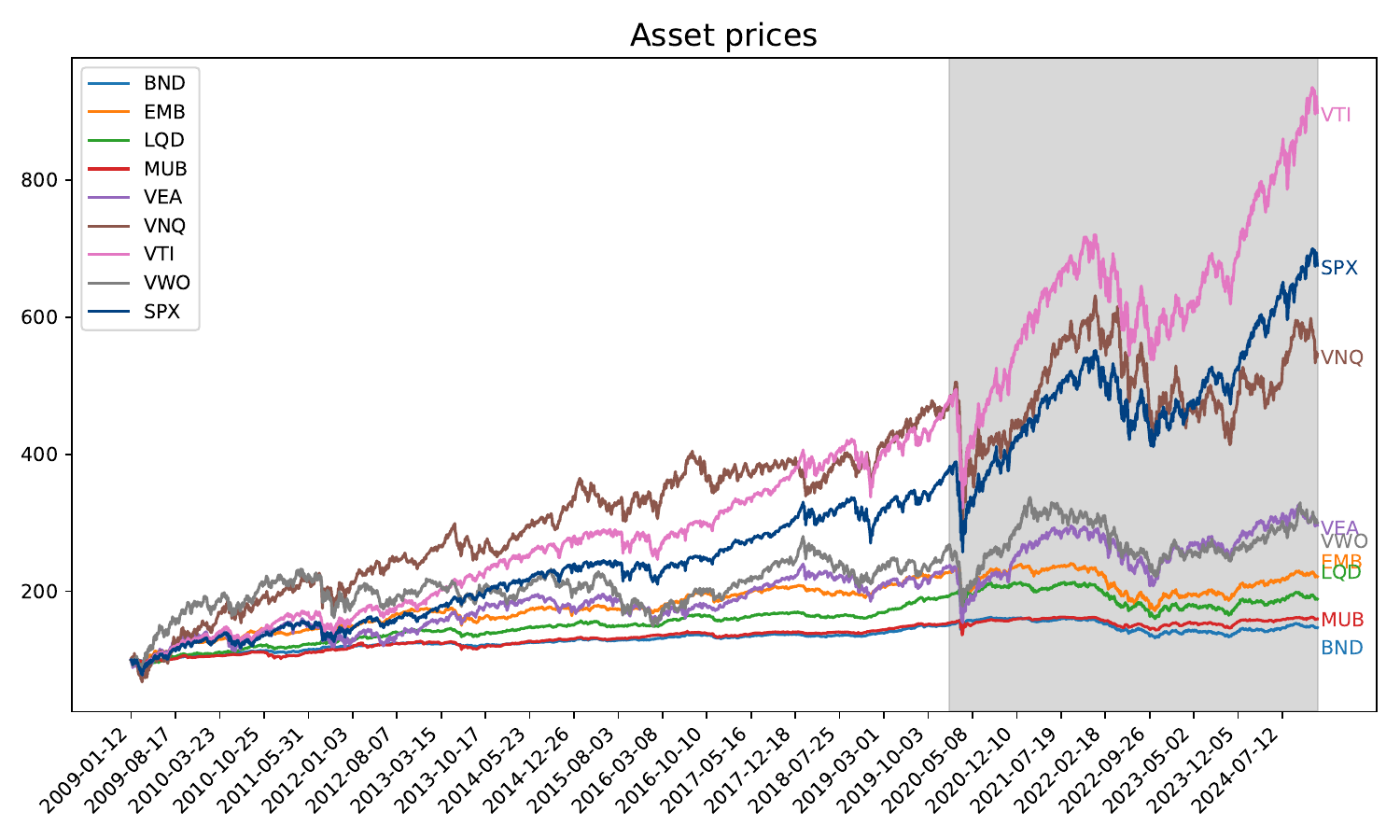}
	\caption{Historical asset prices, scaled to 100 at the first observation. The shaded area represents the trading period, January 16 2020 to December 31, 2024.}
	\label{fig:all-prices}
\end{figure}

{\small
	\begin{table}[h]
		\centering
		\begin{tabular}{lccccccccc}
			\hline
			\rowcolor{lightgray!30}		& BND & EMB & LQD & MUB & VEA & VNQ & VTI & VWO & SPX \\
			\hline
			Mean      & 0.025  & 0.049  & 0.044  & 0.032  & 0.084  & 0.131  & 0.151  & 0.088  & 0.133  \\
			St. Dev.  & 0.049  & 0.099  & 0.078  & 0.050  & 0.196  & 0.257  & 0.182  & 0.223  & 0.180  \\
			Median    & 0.059  & 0.093  & 0.106  & 0.046  & 0.185  & 0.208  & 0.207  & 0.182  & 0.179  \\
			75\%  quant.     & 0.424  & 0.708  & 0.654  & 0.326  & 1.646  & 1.859  & 1.531  & 1.944  & 1.467  \\
			25\%  quant.    & -0.344 & -0.569 & -0.542 & -0.238 & -1.339 & -1.525 & -1.009 & -1.745 & -0.978 \\
			\hline
		\end{tabular}
		\caption{Annualized key statistics for  ETFs and S\&P500 returns.}
		\label{tab:key-stats}
	\end{table}
}

{\small
	
	\newcounter{temptable}
	\setcounter{temptable}{\value{table}}
	\newcounter{tempfigure}
	\setcounter{tempfigure}{\value{figure}}
	\begin{figure}[h]
		\centering
		\renewcommand{\figurename}{Table}
		\setcounter{figure}{\value{table}}
		\addtocounter{figure}{0}
		\includegraphics[width=0.9\textwidth]{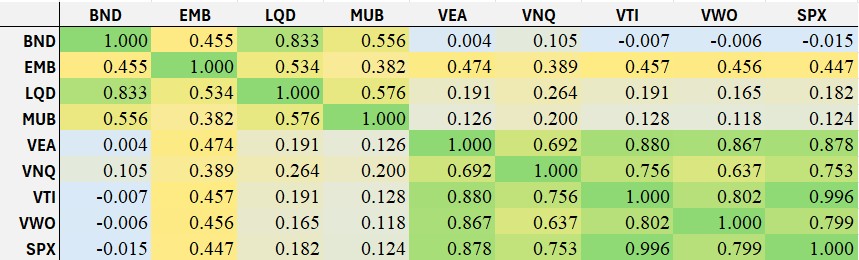}
		\caption{Correlation matrix of asset returns.}
		\label{fig:corr-matrix}
		\renewcommand{\figurename}{Figure} 
		\setcounter{figure}{\value{tempfigure}} 
		\setcounter{table}{\value{figure}} 
		\setcounter{table}{\value{temptable}} 
		\addtocounter{table}{1} 
	\end{figure}
}

\subsection{Computational Methods}
The computational experiments and the development of the RA platform were done using Python 3 (version 3.13.7) programming language, on a PC (Core Ultra 7 165H, 1.40 GHz, with 32 GB RAM, 16 Cores) running Windows~11~Pro.

\begin{figure}[h]
	\centering
	\includegraphics[width=0.8\textwidth]{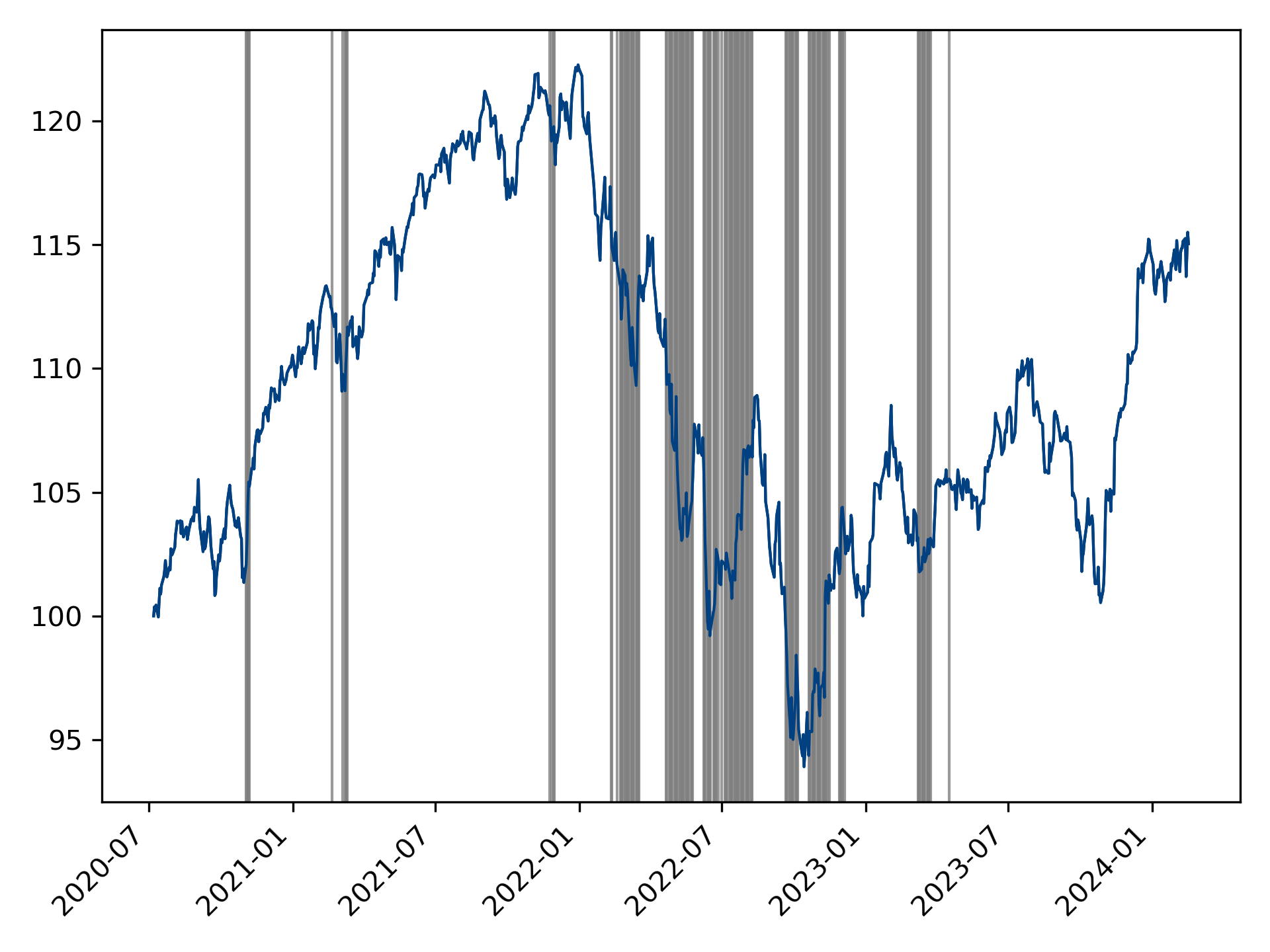}
	\caption{Forecasted market regimes by HMM model, gray shaded area corresponds to `contraction' regime. Dark blue curve represents the average prices of all assets.}
	\label{fig:HMM}
\end{figure}

For estimation and calibration of HMM model, we used \texttt{hmmlearn} (version 0.3.2) library, which uses the Viterbi algorithm to forecast the hidden states and the Baum-Welch algorithm to estimate the model parameters. Specifically, in line with our Gaussian model assumptions, see Section~\ref{sec:forecast}, we use \texttt{hmm.GaussianHMM} class to fit a two state HMM. For this step, we used past five years of historical daily data with rolling window. In Figure~\ref{fig:HMM}, we display the HMM predicted states of the economy. The vertically shaded gray area corresponds to the $c$ regime, namely the contraction economy phase. By visual inspection, we note that indeed sharp declines in the prices where picked up by the algorithm.

We note that all optimization problems are convex, particularly due to the convex approximation for risk budgeting criteria (Section~\ref{sec:SMPC-MRB-Convex}). This allows the use of computationally efficient open-source optimization routines. We use the Python-embedded modeling language \texttt{cvxpy} (version 1.5) for all optimization problems. Thanks to the computational efficiency of the MPC methodology, the computation time for a single strategy is on the order of milliseconds,  making it scalable for managing portfolios across a large number of investors.

\subsection{Set of parameters}

We ran a wide range of combinations of model parameters, within ranges found in existing literature \cite{CostaKwon2018,LI2021,Harris2017,AlonsoSrivastava2022,OprisorKwon2020} but adjusted to our modeling scales:
\begin{itemize}
	\item Number of assets $N=8$;
	\item MPC rolling horizon $H\in\set{1,2,3,5,7,10,15,20}$;
	
	\item Transaction costs sensitivity factor $\eta\in\set{0, 0.0001, 0.001, 0.005, 0.007, 0.01, 0.05, 0.1, 0.5, 1}$,
	\item Turnover constraint bound $\delta\in\set{0.001,  0.003, 0.005, 0.008, 0.01, 0.03, 0.05, 0.08, 0.1, 0.15, +\infty}$;

	\item Risk attitude parameter in MV problem $\gamma\in\set{0.001, 0.01, 0.1, 0.5, 1, 3, 5, 10, 25}$; 

	\item Risk-budgeting parameters:
	\begin{itemize}
		\item sensitivity coefficient $\phi\in\set{0.01, 0.02, 0.05, 0.08, 0.1, 0.15, 0.2, 0.3, 0.5, 0.7, 1}$;
		\item the risk attitude coefficient $\gamma_{R}$ is chosen following Remark~\ref{rem:gammaR}.(i), by dividing the assets in two groups: the risky assets VTI, VEA, VWO, VNQ  and less risky assets	BND, MUB, LQD, EMB. We consider $\gamma_{R}\in\set{0.001, 0.005, 0.01, 0.1, 0.5, 1, 1.5, 2, 5, 10}$.						
		\item initial value $\rho^0=0.6\in(0,1)$ in the recursion computation of $\rho^k$;
		\item regularization coefficient in convexification, $\kappa=0.5\in(0,1)$;
		\item iterative weight coefficient $\xi=0.5\in(0,1)$;
		\item error tolerance $\varepsilon=0.01$;

	\end{itemize}
	\item Black-Litterman:
	\begin{itemize}
		\item average market risk-aversion in two regimes, $\overline{\lambda}_{0}\in\set{0.1, 0.3, 0.5, 1, 1.5, 2, 2.5, 3, 3.5, 4, 5}$ and $\overline{\lambda}_n=1.2\cdot\lambda_0$, $\overline{\lambda}_c=0.8\cdot \lambda_0$;
		\item uncertainty of CAPM as scaling coefficients of the covariance matrix in the priors, $\iota_{BL,n}\in\set{0.01, 0.03, 0.05, 0.1, 0.5, 1}$ and $\iota_{BL,c}=0.9\cdot\iota_{BL,n}$;
		\item confidence in the views, $\alpha_{n}=\alpha_{c}\in\set{0.01, 0.1, 0.5, 1, 5, 10}$.
	\end{itemize}
	
	\item Target portfolio risk attitude coefficient $\gamma_{\textrm{target}} \in \set{0.5,1,2}$; see also Section~\ref{sec:target-portfolio}.
\end{itemize}

We focus our analysis on the following parameters $\gamma$, $\gamma_{RP}$, $\eta$, $\phi$, $H$, $\delta$, and $\gamma_{\textrm{target}}$.
We noticed that values around the specific values of $\rho^0, \kappa, \xi, \varepsilon$ we consider do not significantly change the overall portfolio performance.  Therefore, for tractability, we fix their values. As mentioned earlier, we consider total absolute views with $K=N$, $P_c=P_n=I_{N\times N}$, and the views given by \eqref{BL-view-HMM}.  Additionally, for most of our experiments, we choose  $\alpha_{c/n}=1$, $\overline{\lambda}_0=1$, $\iota_{BL,n}=0.03$. Calibration to these parameters is more of an art. Our choices were based on overall performance of the portfolios and extensive numerical runs.

\subsection{Employed RA strategies and benchmarks portfolios}

\textbf{MV-BL} will denote the RA strategy with mean-variance criteria in the context  of SMPC and using HMM combined with Black-Litterman methodology; see Section~\ref{sec:forecast}, and in particular Remark~\ref{rem:SMPC-HMM-BL}.  Correspondingly, \textbf{MRB-BL} stands for mean-risk-budgeting criterion, with SMPC, HMM and Black-Litterman elements. Along these two main RA strategies developed in this paper, for comparison, we also present the results for some traditional or benchmark strategies, described next.

\textbf{MV-Est-myopic} strategy uses mean-variance criteria, with mean and variance estimated using sample mean and sample variance. The optimization horizon is equal to one rebalancing period,  $H=1$, hence myopic. This is the approach mostly used in financial advising industry. \textbf{MV-Est-MPC}, similar to MV-EST-myopic approach, uses mean-variance criteria with sample mean and variance, but combined with MPC approach by taking a longer horizon $H>1$; see Section~\ref{sec:MV-SMPC}. This approach can be termed as traditional MPC as appearing in engineering contexts.

Additionally, we list the market portfolio \textbf{SP500} equal to the  Standard \& Poor’s 500 Index, as well as the equal weighted portfolio \textbf{1/N}.

We display the dynamics of weights $\pi_\tau$, and the wealth process of the RA strategies scaled to 100 at the starting time $t=0$, January 2, 2020. In all figures, brighter-colored labels indicate riskier assets. Whenever necessary, we also report some annualized key performance metrics, computed using daily  portfolio returns. In particular, we compute the mean return, the standard deviation of the returns (StDev), the maximum drawdown (MaxDD), Sharpe Ratio (SR), Gains-to-Loss ratio (GLR), Calmman Ratio (CR), and turnover. We recall that $\textrm{GLR}(X) = \bE[X]/\bE[X^-]$ and $\textrm{SR}(X)=\bE[X]/\textrm{StDev}(X)$. With $M_t=\max_{0\leq s \leq t} X_t$ denoting the running maximum, the maximum drawdown is defined as 
$\textrm{MaxDD}(X) = \max_{0 \leq t\leq T} (M_t-X_t)/X_t$, and then $\textrm{CR}(X) = \bE[X]/ \textrm{MaxDD}(X)$. Turnover is defined as 
\[
\text{Turnover}(\pi) 
= \frac{1}{2} \cdot 252 \cdot 
\frac{ \displaystyle \sum_{t=2}^{T} \sum_{i=1}^{N} \bigl| \pi_{i,t} - \pi_{i,t-1} \bigr| }{T-1}
\]

\subsection{Analysis of results}
We present a series of results on the sensitivity of the studied RA systems to different parameter sets, structured across several subsections.
Unless otherwise noted, all other constraints and parameters are taken the same for all strategies.

\subsubsection{Risk attitude parameter $\gamma$ in MV criteria}

%

\begin{figure}[h]
	\centering
	\includegraphics[width=0.99\textwidth]{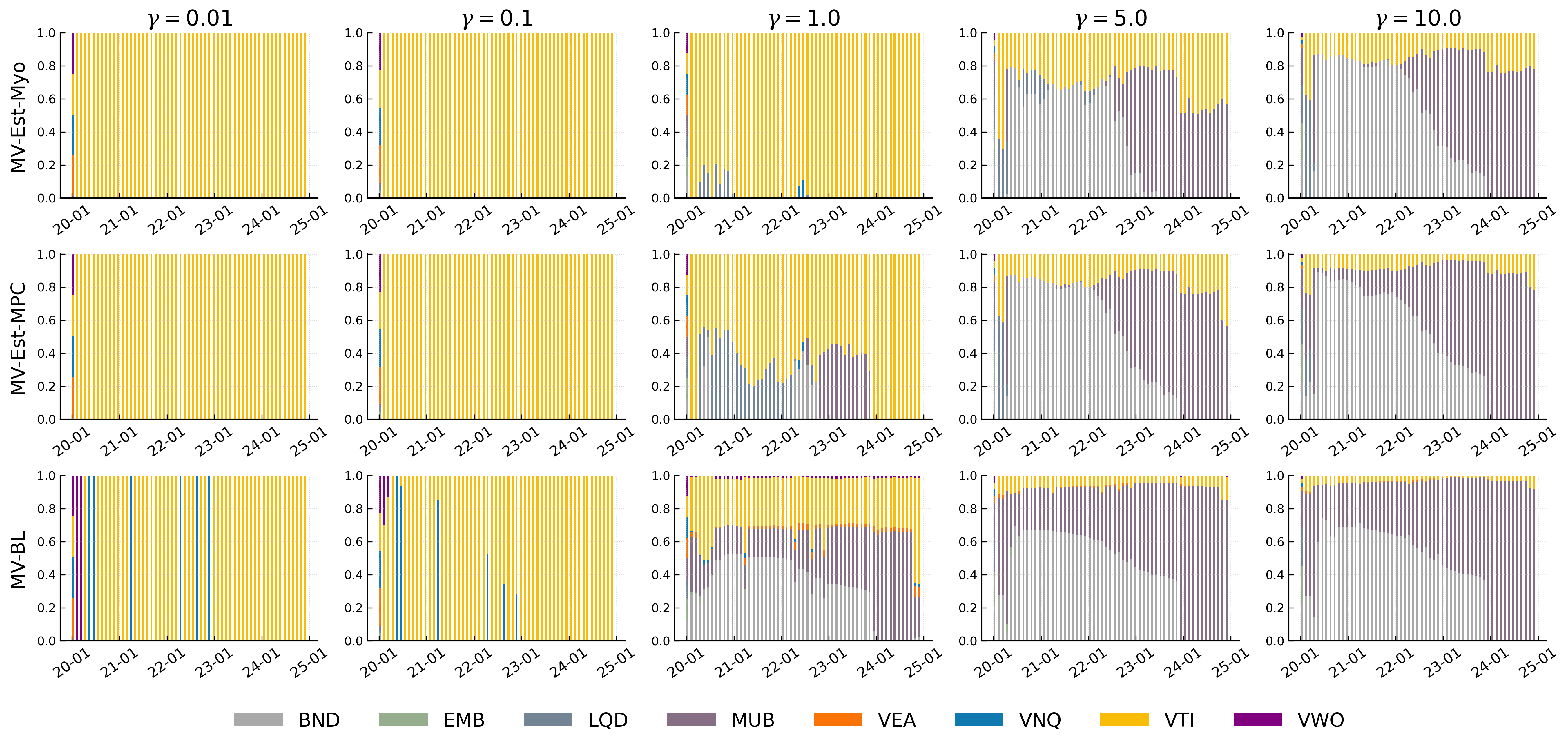}
	\caption{Portfolio weights for  the mean-variance strategies MV-Est-MPC (top row) and MV-BL (bottom row) with risk aversion coefficient  $\gamma\in\set{0.01, 0.1, 1, 5, 10}$.}
	\label{fig:weights-gamma1}
\end{figure}

\begin{figure}[h]
	\centering
	\includegraphics[width=0.99\textwidth]{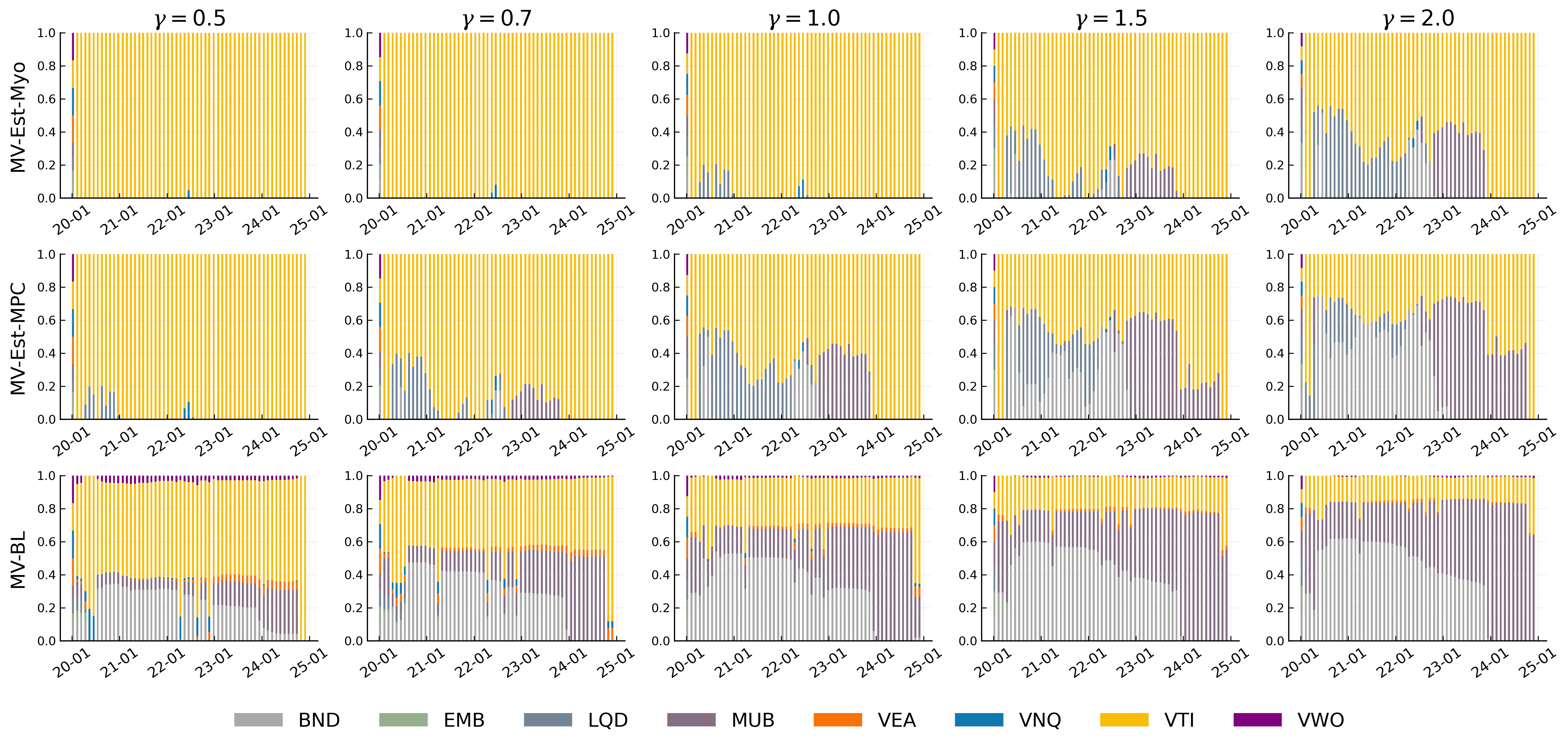}
	\caption{Portfolio weights for  the mean-variance strategies MV-Est-MPC (top row) and MV-BL (bottom row) with risk aversion coefficient $\gamma\in\set{0.5, 0.7, 1, 1.5, 2}$.}
	\label{fig:weights-gamma2}
\end{figure}

\begin{figure}[h]
	\centering
	\includegraphics[width=0.99\textwidth]{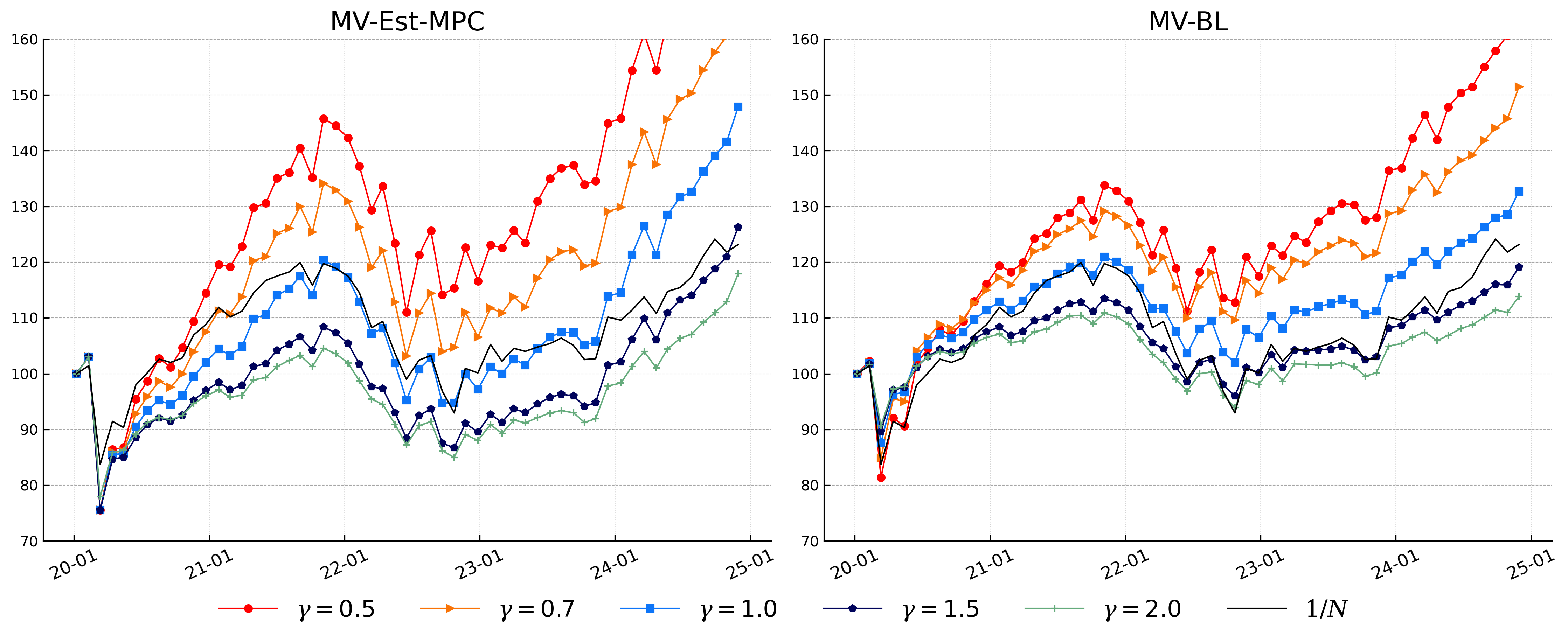}
	\caption{Portfolio wealth for the mean-variance strategies MV-EST-MPC (left panel) and MV-BL (right panel) with $\gamma\in\set{0.5, 0.7, 1, 1.5, 2}$, and for equally weighted portfolio 1/N.}
	\label{fig:wealth-gamma}
\end{figure}

We start with the risk aversion parameter $\gamma$ appearing in the mean-variance criteria. We fix $H=7, \delta = +\infty, \eta=0$.    In Figure~\ref{fig:weights-gamma1} and Figure~\ref{fig:weights-gamma2} we display the portfolios' structure for a wide range of values of parameter $\gamma$ for MV-Est-myopic, MV-Est-MPC and MV-BL strategies, starting with a coarser grid in Figure~\ref{fig:weights-gamma1}, and focusing on $\gamma$ values around one in Figure~\ref{fig:weights-gamma2}, for which we see a shift in portfolio composition for MV-Est-MPC around $\gamma=1.5$, and for MV-BL around $\gamma=1$. As expected, for larger $\gamma$, all three strategies become more conservative, investing mostly in bonds.  We note that the developed strategy MV-BL yields a more diversified portfolio overall, while MV-Est-myopic and MV-Est-MPC exhibit a well-documented drawback of classical mean-variance analysis--namely, investing in a few assets, with a sharp transition between being too conservative (all in cash or a few bonds) and too risky (all in a few risky assets). In Figure~\ref{fig:wealth-gamma} we display the wealth process for these strategies, along with the equally weighted strategy 1/N. Key performance metrics are presented in Table~\ref{tab:gammas}. As expected, smaller $\gamma$ (less risk averse investing profile), yields larger returns but also larger standard deviations of the returns. The Sharpe Ratio and Gain-to-Loss ratio both decrease as $\gamma$ increases.

The myopic strategy, MV-Est-myopic, exhibits the least desirable behavior from a robo-advising perspective, shifting from an ``all-in-one'' risky asset allocation to just a few assets within a narrow range of the risk-tolerance parameter $\gamma$. This behavior was consistently observed across other parameter sets. Consequently, we exclude this strategy from the analysis presented below and do not recommend it for any robo-advisor aiming to implement a robust and comprehensive system.

{\small
	\begin{table}[h!]
		\centering
		\renewcommand{\arraystretch}{1.0} 
		\begin{tabular}{rcccccccc}
		\hline
			\rowcolor{lightgray!30}		&  Mean & StDev & MaxDD & SR & GLR & CalmR & Turnover \\
		\hline
			\multicolumn{7}{c}{$\gamma=0.5$}  \\		
			M-BL & 0.121 & 0.166 & 0.279 & 0.724 & 0.151 & 0.431 & 0.887 \\
			MV-Est-myopic & 0.161 & 0.217 & 0.350 & 0.743 & 0.137 & 0.461 & 0.190 \\
			MV-Est-MPC & 0.152 & 0.210 & 0.344 & 0.723 & 0.135 & 0.441 & 0.416 \\
			1/N & 0.051 & 0.129 & 0.240 & 0.398 & 0.094 & 0.214 & -- \\
			SP500 & 0.150 & 0.215 & 0.339 & 0.701 & 0.137 & 0.443 & -- \\			
		\hline
			\multicolumn{7}{c}{$\gamma=0.7$} \\		
			M-BL & 0.095 & 0.137 & 0.234 & 0.693 & 0.155 & 0.405 & 0.988 \\
			MV-Est-myopic & 0.161 & 0.217 & 0.350 & 0.742 & 0.136 & 0.460 & 0.207 \\
			MV-Est-MPC & 0.124 & 0.189 & 0.334 & 0.656 & 0.130 & 0.372 & 0.845 \\
		\hline
			\multicolumn{7}{c}{$\gamma=1$} \\		
			M-BL & 0.065 & 0.106 & 0.188 & 0.615 & 0.141 & 0.346 & 0.733 \\
			MV-Est-myopic & 0.152 & 0.210 & 0.344 & 0.723 & 0.135 & 0.441 & 0.431 \\
			MV-Est-MPC & 0.093 & 0.161 & 0.336 & 0.580 & 0.129 & 0.278 & 1.134 \\
		\hline
			\multicolumn{7}{c}{$\gamma=1.5$} \\		
			M-BL & 0.040 & 0.085 & 0.165 & 0.468 & 0.115 & 0.240 & 0.644 \\
			MV-Est-myopic & 0.116 & 0.183 & 0.334 & 0.635 & 0.128 & 0.348 & 0.979 \\
			MV-Est-MPC & 0.058 & 0.138 & 0.337 & 0.418 & 0.106 & 0.171 & 1.501 \\
		\hline
			\multicolumn{7}{c}{$\gamma=2$} \\		
			M-BL & 0.030 & 0.078 & 0.157 & 0.381 & 0.098 & 0.190 & 0.569 \\
			MV-Est-myopic & 0.093 & 0.161 & 0.336 & 0.577 & 0.128 & 0.277 & 1.155 \\
			MV-Est-MPC & 0.041 & 0.119 & 0.310 & 0.345 & 0.093 & 0.133 & 1.299 \\
		\hline
		\end{tabular}
		\caption{Performance metrics for the MV strategies with various $\gamma$. }
		\label{tab:gammas}
	\end{table}
}

\subsubsection{Risk attitude parameter $\gamma_{R}$ in MRB and impact coefficient $\phi$}

\begin{figure}[h]
	\centering
	\includegraphics[width=0.99\textwidth]{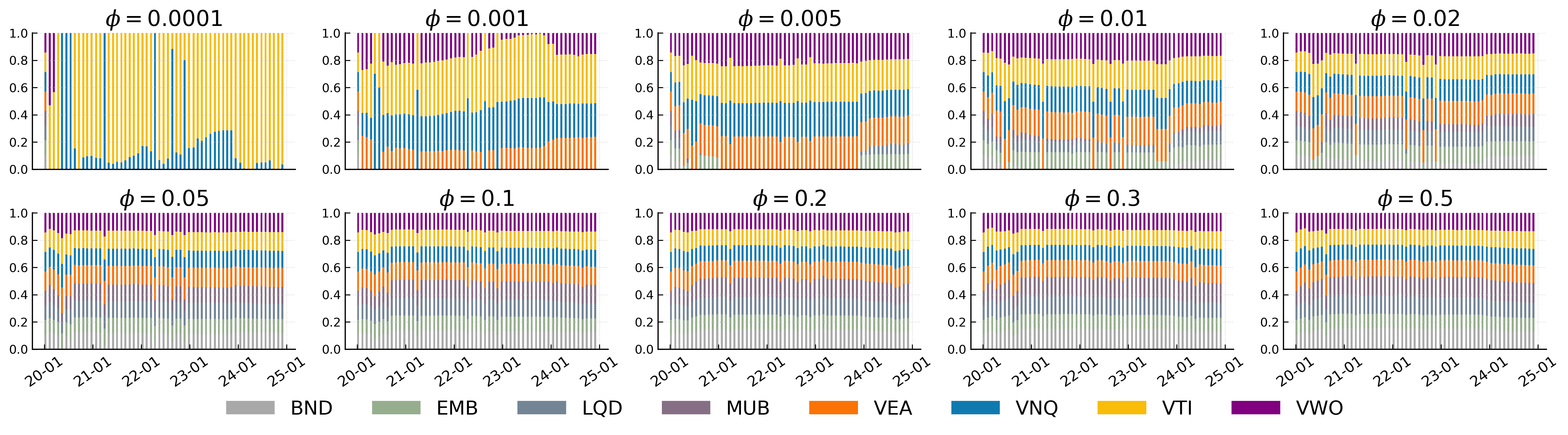}
	\caption{Portfolio weights for the MRB-BL  for various sensitivity coefficient $\phi$.}
	\label{fig:weights-phi}
\end{figure}

The sensitivity coefficient $\phi$ aims to capture the level at which the penalty term in the MRB is enforced. In Figure~\ref{fig:weights-phi} we present the optimal portfolio weights for a series of values of $\phi$, and fixed $\gamma_R=0.5$. While there is some variability in the portfolio composition for smaller values of $\phi$, the results stabilize for $\phi >0.05$. In what follows, we take $\phi=0.1$ or $\phi=0.05$.

\begin{figure}[h]
	\centering
	\includegraphics[width=0.99\textwidth]{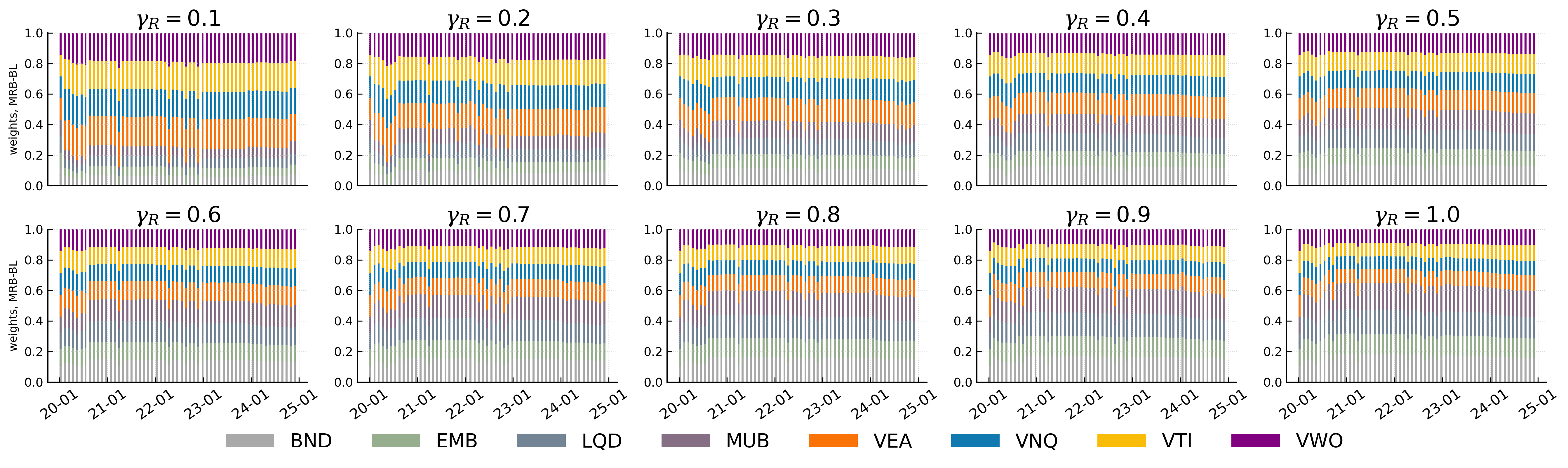}
	\caption{Portfolio weights for the MRB-BL strategy for risk attitude coefficient  $\gamma_R\in\set{0.1, 0.2, 0.3, 0.4, 0.5, 0.6, 0.7, 0.8, 0.9, 1}$.}
	\label{fig:weights-gammaR2}
\end{figure}

\begin{figure}[h]
	\centering
	\includegraphics[width=0.99\textwidth]{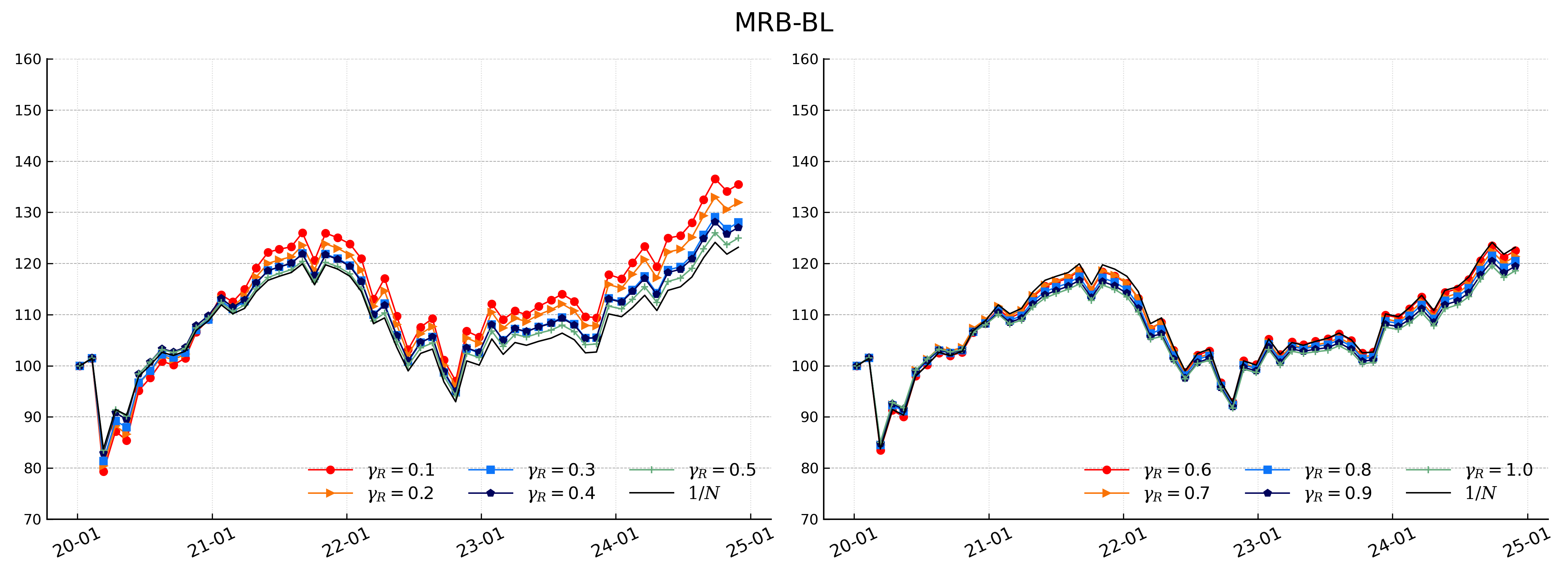}
	\caption{Portfolio wealth for the MRB-BL strategy  with $\gamma_R\in\set{0.5, 0.7, 1.5, 2}$, and for equally weighted portfolio 1/N.}
	\label{fig:wealth-gammaR}
\end{figure}

We recall that in the MRB-BL strategy the parameter $\gamma_R$ plays the role of a risk attitude  coefficient, by shifting the risk in the weights $b_i$  appearing the MRB penalty term towards bonds with increasing values of $\gamma_R$.   This characteristic feature is confirmed in Figure~\ref{fig:weights-gamma1} for a wide range of values $\gamma_R\in\set{0.001, 0.005, 0.01, 0.1, 0.05, 1, 1.5, 2, 5, 10}$. By visual inspection, we note that the portfolios' composition varies for $\gamma_R$ in the range $(0.01,1)$, while remains stable for smaller or larger values. Optimal portfolio weights for $\gamma_R$ on a finer grid around 0.5 are presented in Figure~\ref{fig:weights-gammaR2}, along the wealth process of these portfolio shown in Figure~\ref{fig:wealth-gammaR}. Key performance metrics are presented in Table~\ref{tab:gammasR}. Sharpe Ratio and GLR are decreasing in $\gamma_R$. Calman Ratio also decreases with few exceptions.

Similar to the MV strategies, portfolio wealth decreases as $\gamma_R$ increases, reflecting higher investor risk aversion. However, the magnitude of these changes is substantially smaller. We further emphasize that the optimal weights $\pi_\tau$ under MRB–BL exhibit significantly less time variation compared to MV–BL. This is a distinctive feature of the method, stemming from the fact that the risk profiles of individual assets remain relatively stable over the investment horizon and that risk allocation in MRB is enforced at the individual-asset level. Consequently, for a fixed $\gamma_R$, the portfolio structure remains largely stable over time.

From a robo-advising perspective, this relative stability has mixed implications. On the one hand, slowly varying portfolio weights may reduce portfolio turnover or transaction costs. We argue that these type of properties are better incorporated directly through the optimization criterion or via hard constraints; see Section~\ref{sec:tr_costs} and~\ref{sec:numerical-eta-delta}. Moreover, reduced sensitivity to changing forecasts may limit the responsiveness of the allocation to evolving market conditions. These considerations suggest that, while MRB-SMPC strategy exhibit appealing structural properties, they should be combined with other mechanism when deployed in practical robo-advising systems.

\begin{figure}[h]
	\centering
	\includegraphics[width=0.99\textwidth]{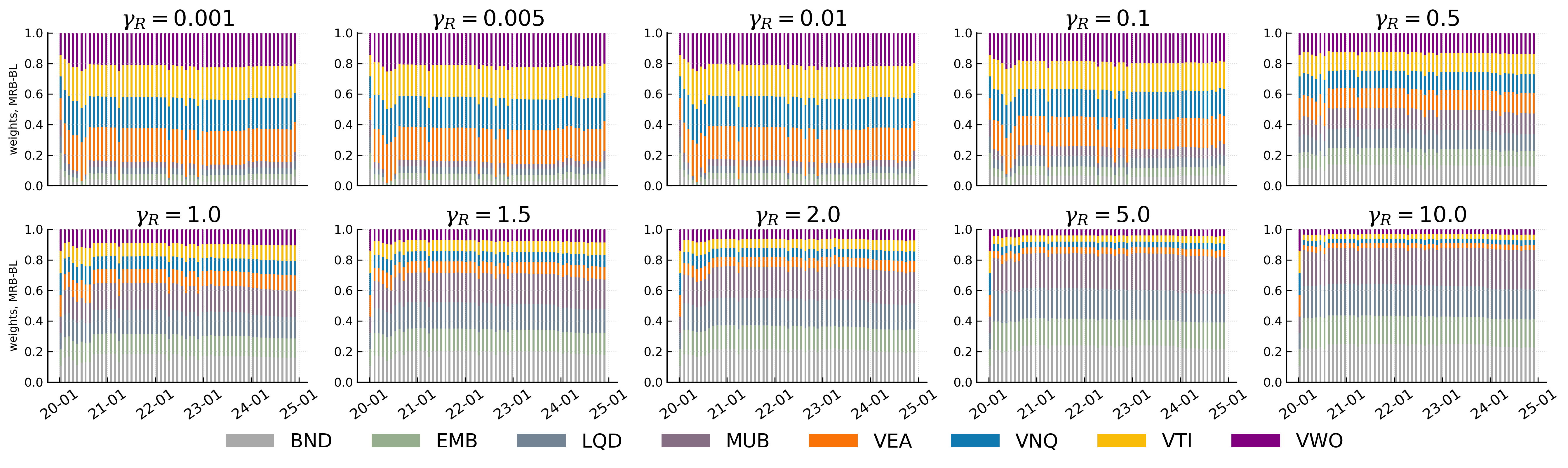}
	\caption{Portfolio weights for the MRB-BL strategy for risk attitude coefficient $\gamma_R\in\set{0.001, 0.005, 0.01, 0.1, 0.5, 1, 1.5, 2, 5, 10}$.}
	\label{fig:weights-gammaR1}
\end{figure}

{\small
	\begin{table}[h!]
		\centering
		\renewcommand{\arraystretch}{1.0} 
		\begin{tabular}{r|cccccccc}
			\hline
			\rowcolor{lightgray!30}	$\gamma_R$	&  Mean & StDev & MaxDD & SR & GLR & CalmR & Turnover \\
			\hline			
0.001 & 0.087 & 0.184 & 0.327 & 0.471 & 0.107 & 0.266 & 0.311 \\
0.005 & 0.086 & 0.184 & 0.325 & 0.466 & 0.106 & 0.264 & 0.321 \\
0.01 & 0.085 & 0.184 & 0.328 & 0.459 & 0.106 & 0.258 & 0.334 \\
0.1 & 0.077 & 0.169 & 0.306 & 0.455 & 0.105 & 0.251 & 0.299 \\
0.3 & 0.062 & 0.146 & 0.276 & 0.422 & 0.099 & 0.223 & 0.222 \\
0.5 & 0.055 & 0.133 & 0.248 & 0.413 & 0.098 & 0.222 & 0.176 \\
0.7 & 0.048 & 0.126 & 0.233 & 0.382 & 0.093 & 0.206 & 0.270 \\
0.9 & 0.044 & 0.119 & 0.229 & 0.368 & 0.091 & 0.192 & 0.289 \\
1 & 0.045 & 0.115 & 0.216 & 0.394 & 0.096 & 0.211 & 0.239 \\
1.5 & 0.036 & 0.106 & 0.210 & 0.337 & 0.082 & 0.170 & 0.218 \\
2 & 0.032 & 0.101 & 0.208 & 0.316 & 0.077 & 0.154 & 0.181 \\
5 & 0.024 & 0.092 & 0.205 & 0.257 & 0.063 & 0.115 & 0.182 \\
10 & 0.020 & 0.087 & 0.203 & 0.223 & 0.056 & 0.096 & 0.170 \\
			\hline
		\end{tabular}
		\caption{Performance metrics for the MRB-BL strategy with various $\gamma_R$. }
		\label{tab:gammasR}
	\end{table}
}

\subsubsection{Transaction cost coefficient $\eta$ and turnover constraint bound $\delta$}\label{sec:numerical-eta-delta}


We recall that both the transaction cost penalty term $\eta \cdot TC(z)$ in the criteria and the turnover constraint $\norm{\pi_{\tau+1}-\pi_\tau}_1\leq \delta$ are
meant to control the magnitude of the overall transaction costs; see Section~\ref{sec:tr_costs}.

\begin{figure}[h]
	\centering
	\includegraphics[width=0.99\textwidth]{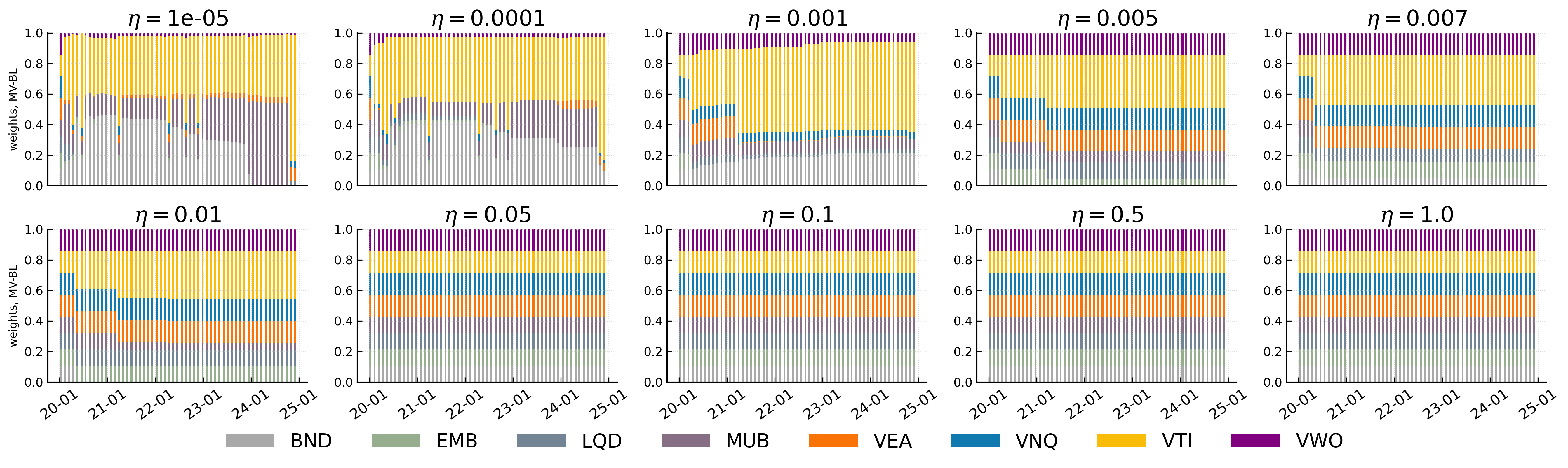}
	\caption{Portfolio weights for the MV-BL strategy  with \\  $\eta\in\set{0, 10^{-6}, 10^{-4}, 0.001, 0.003, 0.005, 0.007, 0.01, 0.03, 0.05}$.}
	\label{fig:weights-etaMV}
\end{figure}

\begin{figure}[h]
	\centering
	\includegraphics[width=0.45\textwidth]{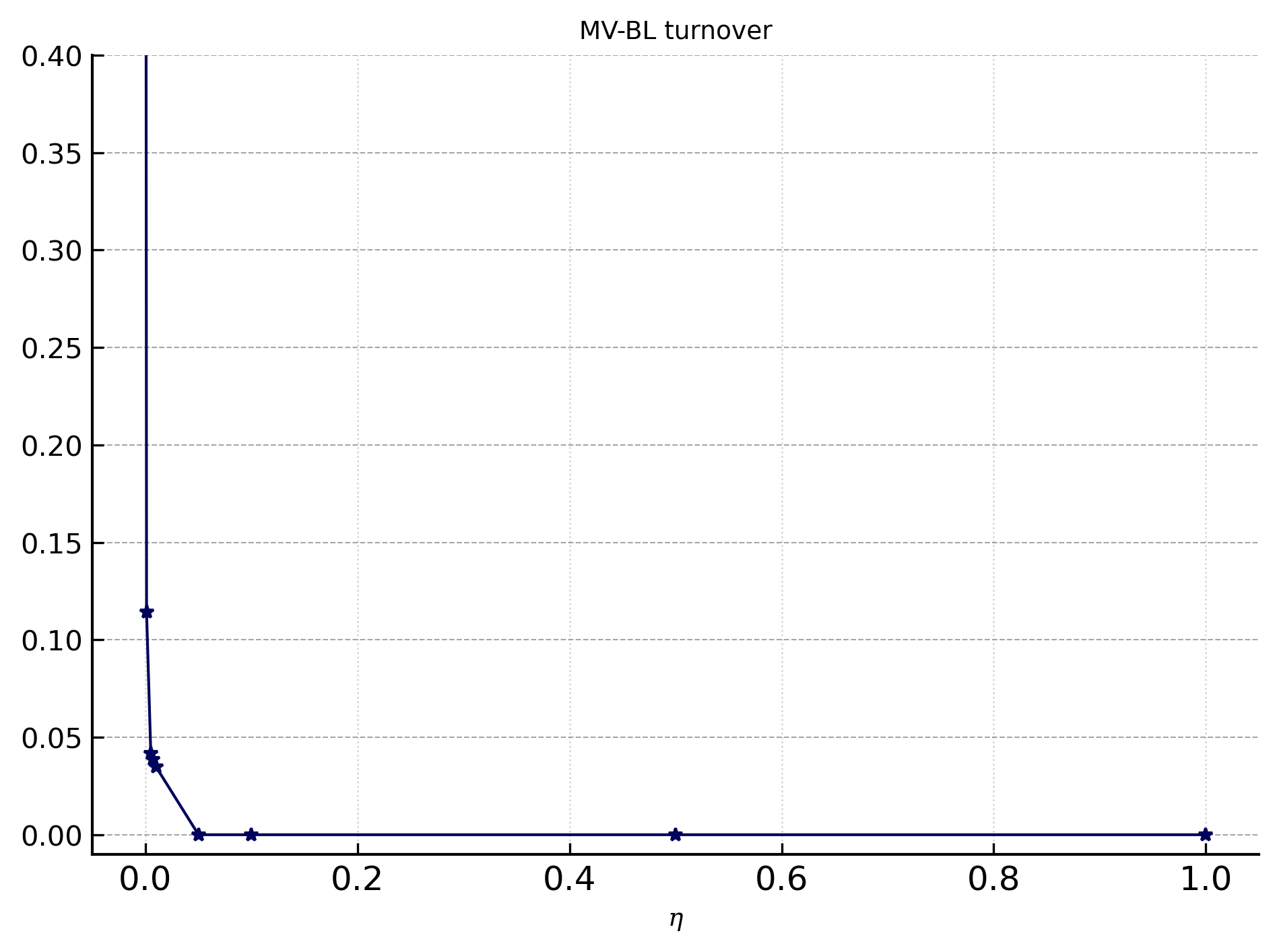} \ \
	\includegraphics[width=0.45\textwidth]{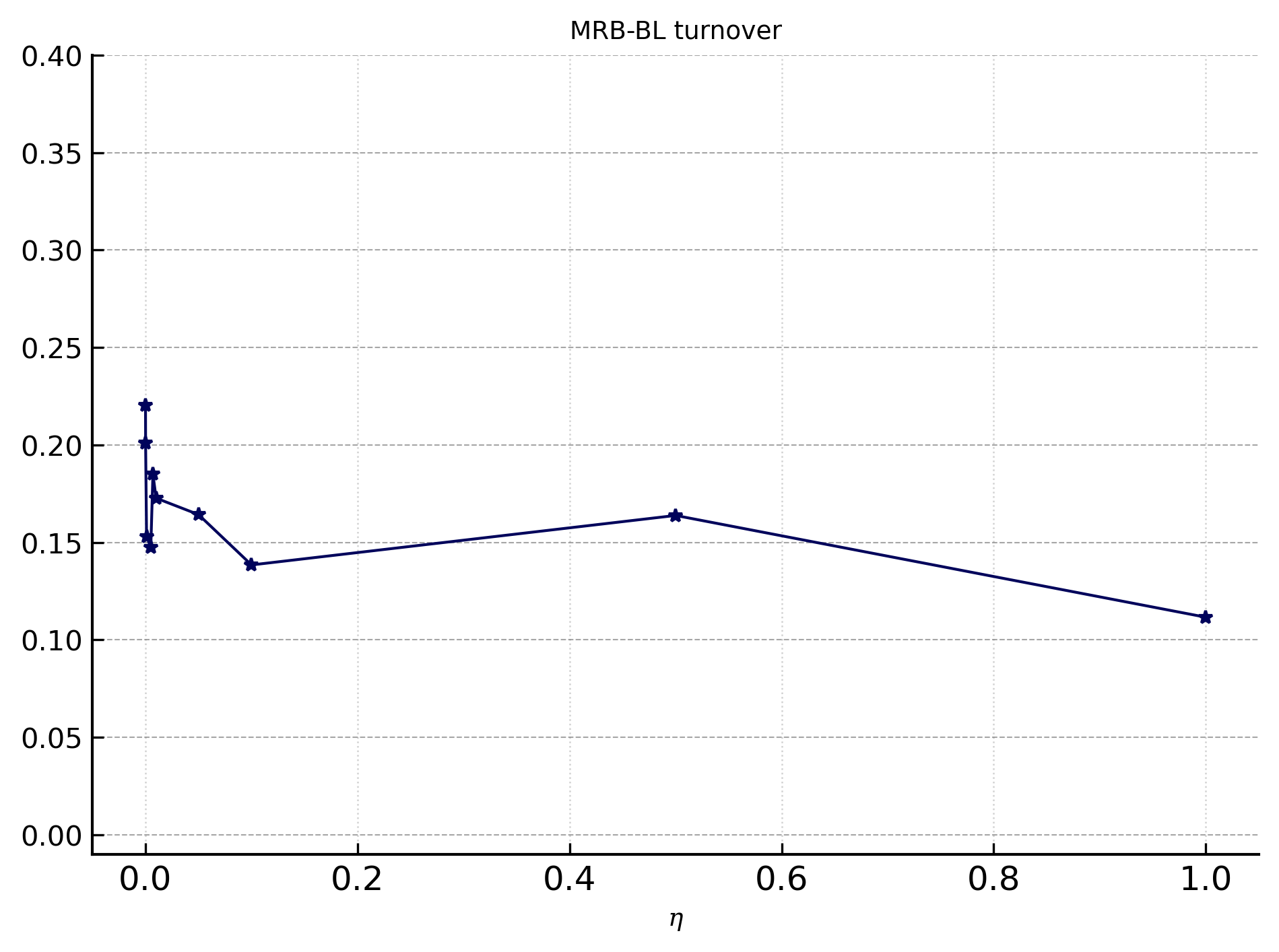}
	\caption{Turnover for MV-BL (left) and MRB-BL (righ) as function of $\eta$.}
	\label{fig:turnover-etaMV}
\end{figure}

We start by analyzing the effect of the transaction cost scaling factor $\eta$ in the criteria, while keeping $\delta=+\infty$. In Figure~\ref{fig:weights-etaMV}, we present the portfolio weights for different values of $\eta$, while in Figure~\ref{fig:turnover-etaMV}, left panel, we plot the total turnover for this strategy as function of $\eta$.
With increasing $\eta$, the portfolio structure changes less in time, and the total turnover decreases and vanishing for $\eta>0.05$.

\begin{figure}[h]
	\centering
	\includegraphics[width=0.99\textwidth]{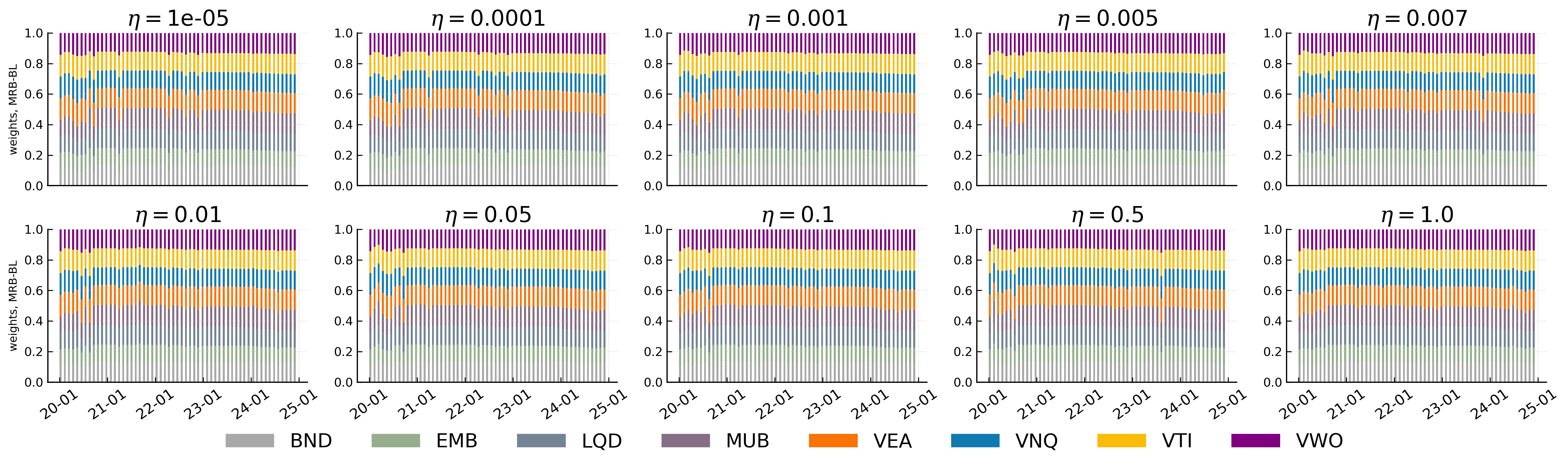}
	\caption{Portfolio weights for the MRB-BL strategy   with \\  $\eta\in\set{0, 0.1, 1,5,10,50,100,500,1000,2000}$.}
	\label{fig:weights-etaMRB}
\end{figure}

On the other hand, as illustrated in Figure~\ref{fig:weights-etaMRB}, the portfolio weights under the MRB-BL strategy--and the corresponding turnover shown in the right panel of Figure~\ref{fig:turnover-etaMV}--remain essentially unchanged as $\eta$ varies.

\begin{figure}[h]
	\centering
	\includegraphics[width=0.99\textwidth]{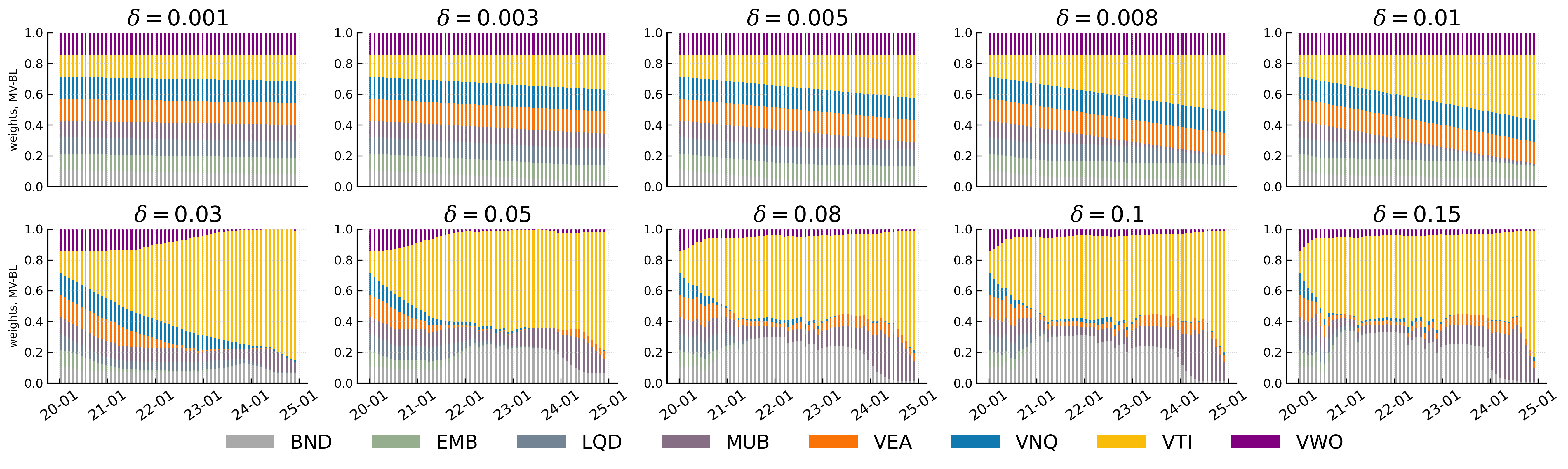}
	\caption{Portfolio weights for the mean-variance strategy MV-BL  with \\  $\delta \in\set{0.001, 0.003, 0.005, 0.008, 0.01, 0.03, 0.05, 0.08, 0.1, 0.15}$.}
	\label{fig:weights-deltaMV}
\end{figure}

\begin{figure}[h]
	\centering
	\includegraphics[width=0.99\textwidth]{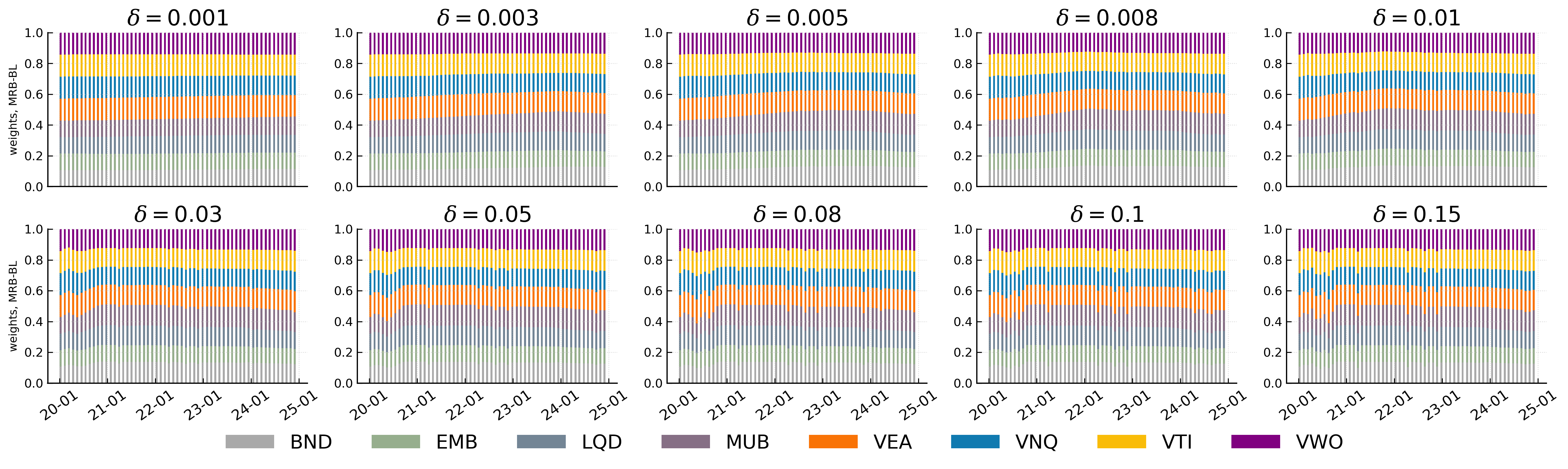}
	\caption{Portfolio weights for the mean-risk-budgeting strategy MRB-BL  with \\  $\delta \in\set{0.001, 0.003, 0.005, 0.008, 0.01, 0.03, 0.05, 0.08, 0.1, 0.15}$.}
	\label{fig:weights-deltaRP}
\end{figure}

Next, we examine sensitivity with respect to $\delta$, while fixing $\eta=0$. In both strategies, the profile is similar: as $\delta$ decreases, turnover declines and the portfolio gradually converges to a static allocation. Conversely, as $\delta$ increases, the portfolio in both cases progressively approaches the unconstrained allocation; see Figures~\ref{fig:weights-deltaMV}--\ref{fig:weights-deltaRP}.

In the context of robo-advising, these results suggest that for the MV-BL strategy, either soft transaction cost penalties or turnover constraints can be used effectively to control trading activity, as both approaches lead to similar portfolio behavior. In contrast, for the MRB-BL strategy, soft constraints are largely ineffective, and one must rely on hard constraints on turnover. This highlights the importance of selecting constraint types that are compatible with the underlying optimization objective when designing automated portfolio management rules.

\begin{figure}[h]
	\centering
	\includegraphics[width=0.45\textwidth]{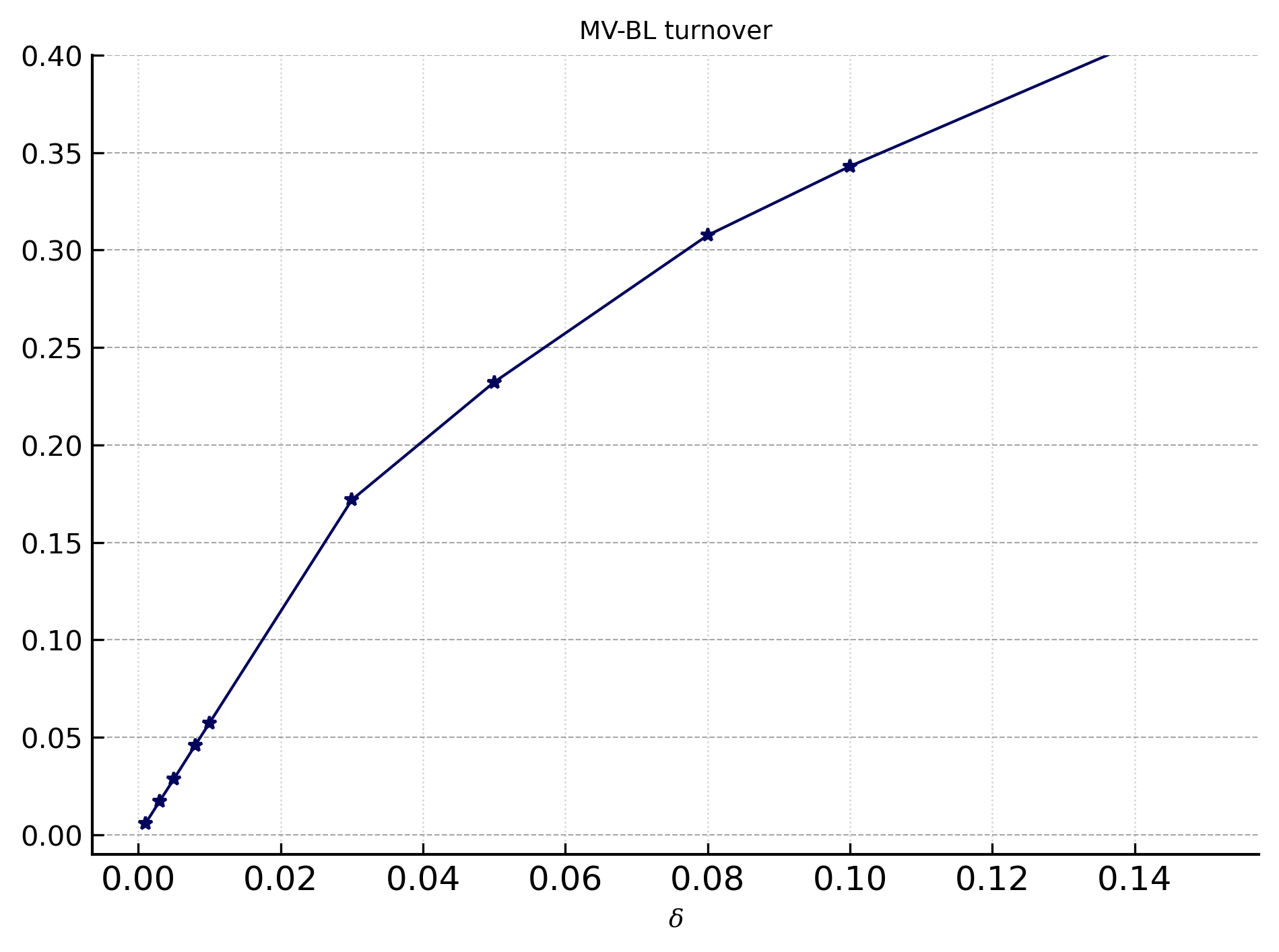} \ \
	\includegraphics[width=0.45\textwidth]{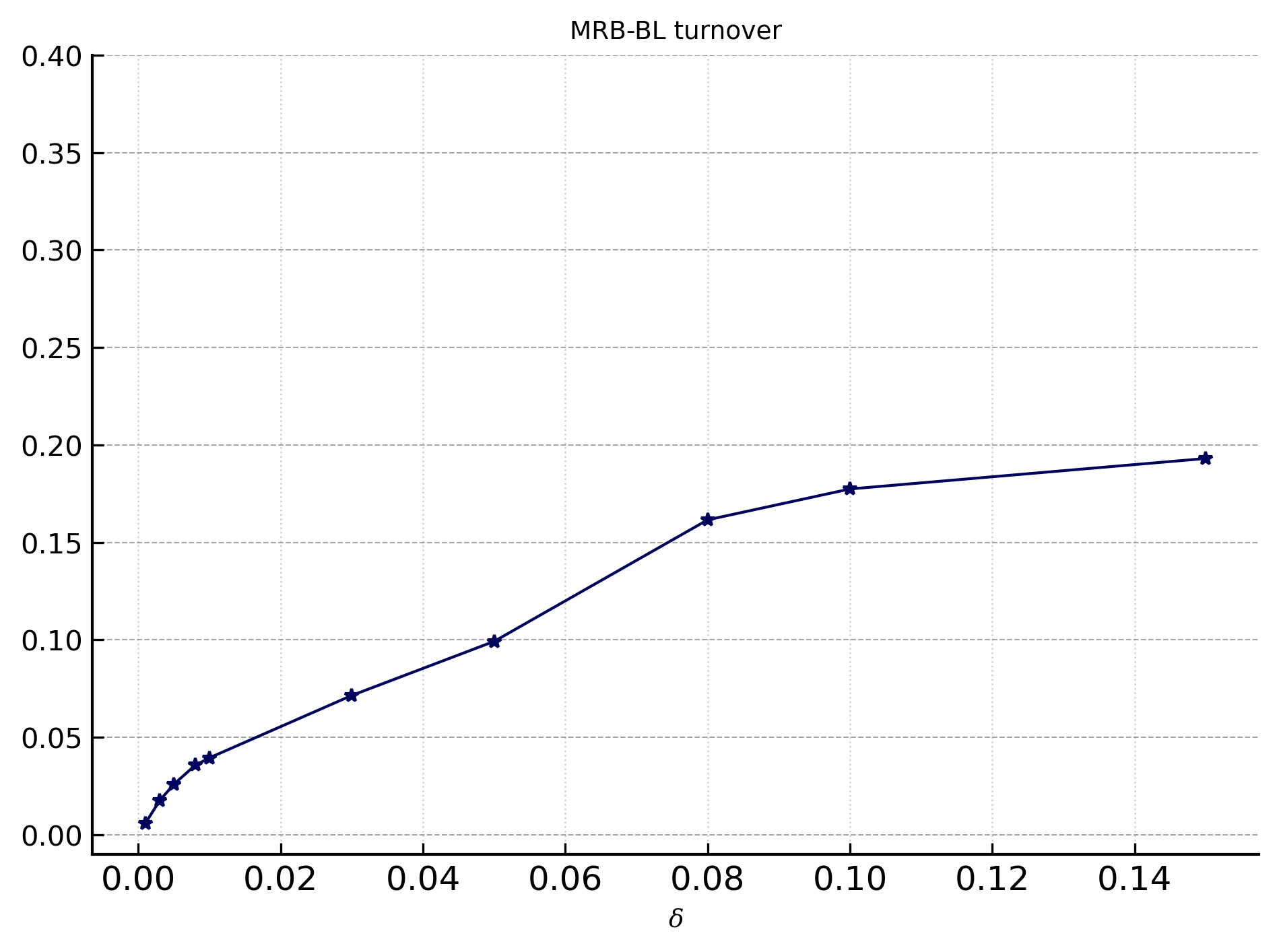}
	\caption{Turnover for MV-BL (left) and MRB-BL (righ) as function of $\delta$.}
	\label{fig:turnover-deltaMV}
\end{figure}

\subsubsection{SMPC rolling horizon parameter $H$}


We now turn our attention to the rolling horizon parameter $H$—one of the central components of the SMPC methodology—which is frequently claimed in applied studies to play a critical role in the performance of the underlying controlled system.

\begin{figure}[h]
	\centering
	\includegraphics[width=0.99\textwidth]{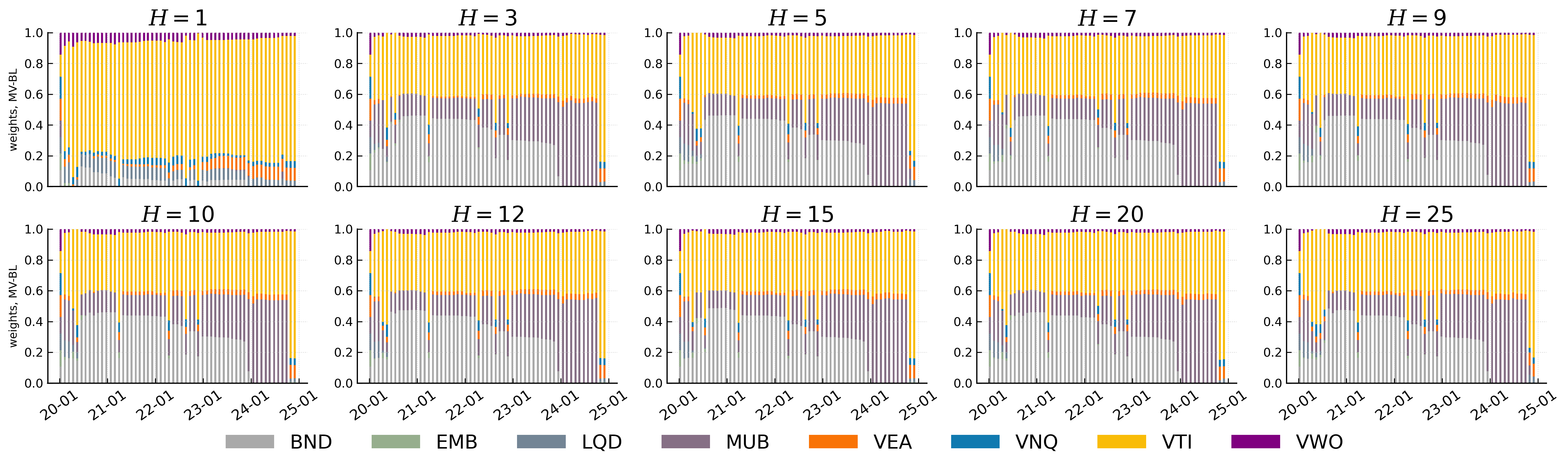}
	\includegraphics[width=0.99\textwidth]{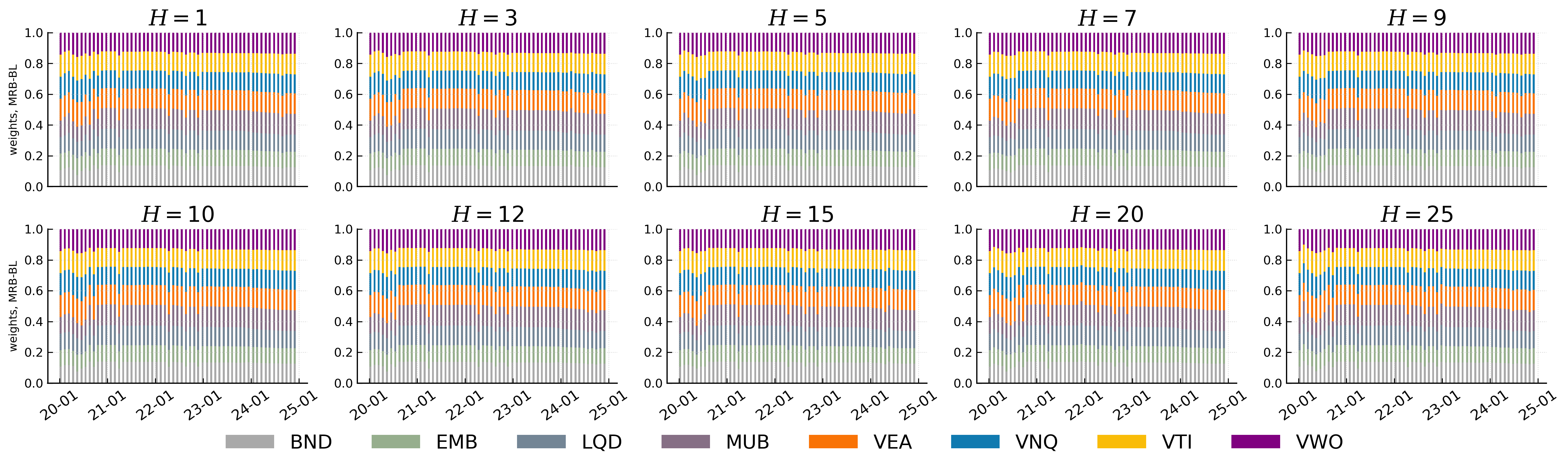}
	\caption{Portfolio weights for MV-BL (top two rows) and MRB-BL (bottom two rows) for $H = 1, 3, 5, 7, 9, 10, 12, 15, 20, 25.$}
	\label{fig:weights-H}
\end{figure}

In Figure~\ref{fig:weights-H}, we display the portfolio weights for the MV-BL and MRB-BL strategies for values of $H$ ranging from 1 to 25, with $\gamma = 0.75$, $\gamma_{R} = 0.5$, $\phi = 0.1$, $\delta = \infty$, and $\eta = 0$. Similar outcomes were consistently observed for other parameter configurations of $\gamma$, $\gamma_R$, $\delta$, $\eta$, and $\phi$. For the MV-BL strategy, performance is stable and optimized for $H$ in the range 5--7, with no meaningful improvements for $H>7$.

For larger values of $H$, the optimization problem at each rebalancing step involves more decision variables and therefore requires additional computation time. While this increase in complexity is negligible for a single portfolio, it may become relevant when deploying the strategy across a large number of portfolios, as is typically the case in large-scale robo-advisory systems. For this reason, we fix $H=7$ as the default value in all subsequent experiments.

In contrast, the performance of the MRB-BL strategy remains essentially unchanged across all considered values of $H$, including the myopic case $H=1$. This indicates that, while the MPC framework combined with HMM and BL forecasts enhances performance under the MV criterion, it provides limited additional benefit for the MRB objective.

\subsubsection{Time changing risk attitude parameters $\gamma$ and $\gamma_R$}\label{sec:num-change-gamma}

%

\begin{figure}[h]
	\centering
	\includegraphics[width=0.99\textwidth]{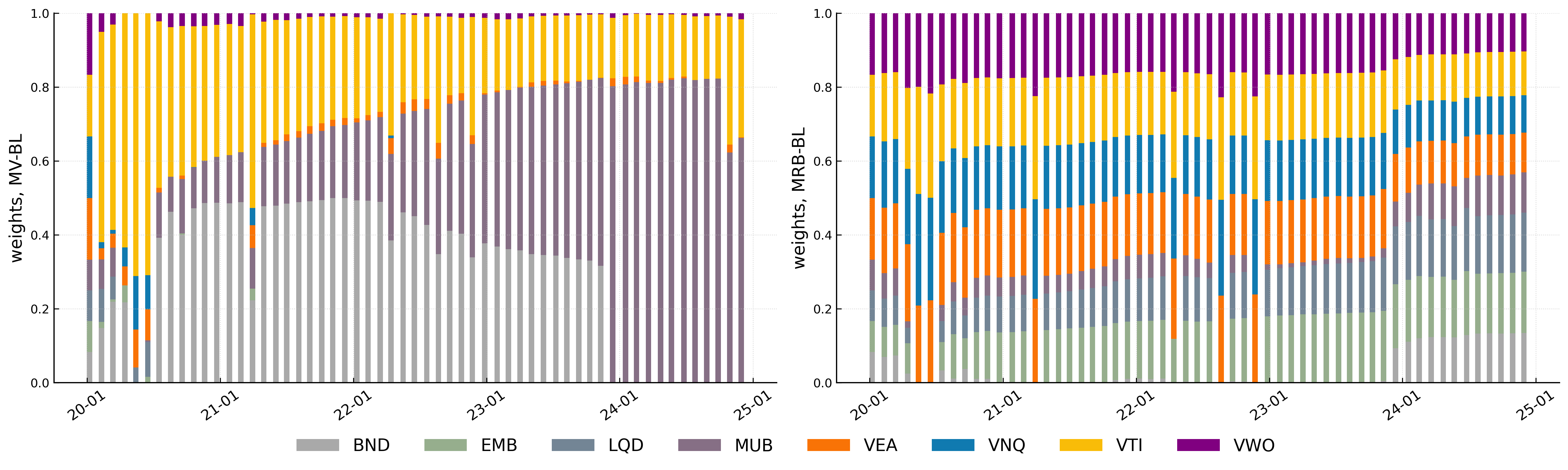}
	\caption{Portfolio weights for the MV-BL and MRB-BL strategies under a linearly time-varying risk attitude parameter $\gamma$.}
	\label{fig:linear-gamma}
\end{figure}

In this section we illustrate how time varying risk attitude parameter affects the portfolios' weights; see also Section~\ref{sec:risk-profiling}. We take $\phi= 0 .01, \eta=0, \delta = +\infty$.
Static profiles, with $\gamma_t, \gamma_{R,t}$ constant in time, have been displayed in the previous section. The lifecyle profile starts with risk tolerance parameters $\gamma_0=0.5$ and $\gamma_{R,0}=0.1$ and increases them linearly to $\gamma_{T-1}=\gamma_{R,T-1}=2$. The portfolios' dynamics are shown in Figure~\ref{fig:linear-gamma}. As expected, in both cases the porfolios shift toward a more risk averse one. Notably, MV-BL just increases the weights of one MUB while decreasing the holding of VTI, visably linearly, while MRB-BL rembalences all assets, putting more weights on all four bonds.

In this section, we illustrate how time-varying risk attitude parameters affect portfolio weights; see also Section~\ref{sec:risk-profiling}. Throughout this experiment, we set $\phi = 0.01$, $\eta = 0$, and $\delta = +\infty$. Static profiles, in which $\gamma_t$ and $\gamma_{R,t}$ are constant over time, were analyzed in the previous section.

The lifecycle profile starts with risk tolerance parameters $\gamma_0 = 0.5$ and $\gamma_{R,0} = 0.1$, which are then increased linearly to $\gamma_{T-1} = \gamma_{R,T-1} = 2$. The resulting portfolio dynamics are shown in Figure~\ref{fig:linear-gamma}. As expected, both strategies shift toward more risk-averse allocations over time. Notably, the MV-BL strategy adjusts primarily along a single dimension, gradually increasing the allocation to MUB while correspondingly reducing exposure to VTI in an almost linear fashion. In contrast, the MRB-BL strategy actively rebalances across all assets, systematically increasing allocations to all bond instruments.

\begin{figure}[h]
	\centering
	\includegraphics[width=0.99\textwidth]{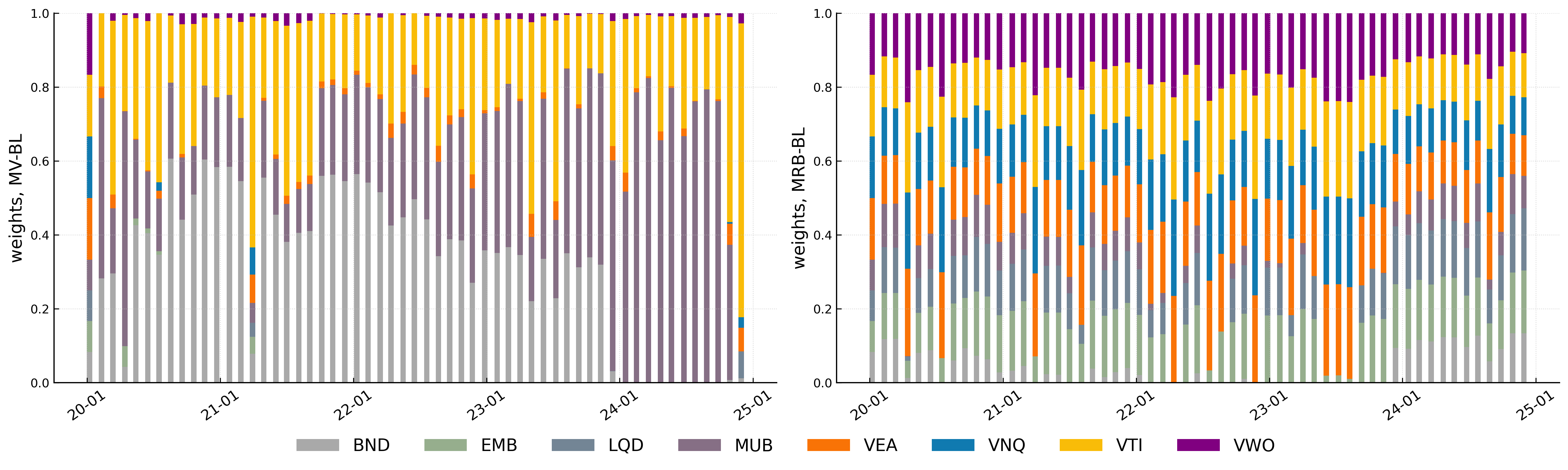}
	\includegraphics[width=0.99\textwidth]{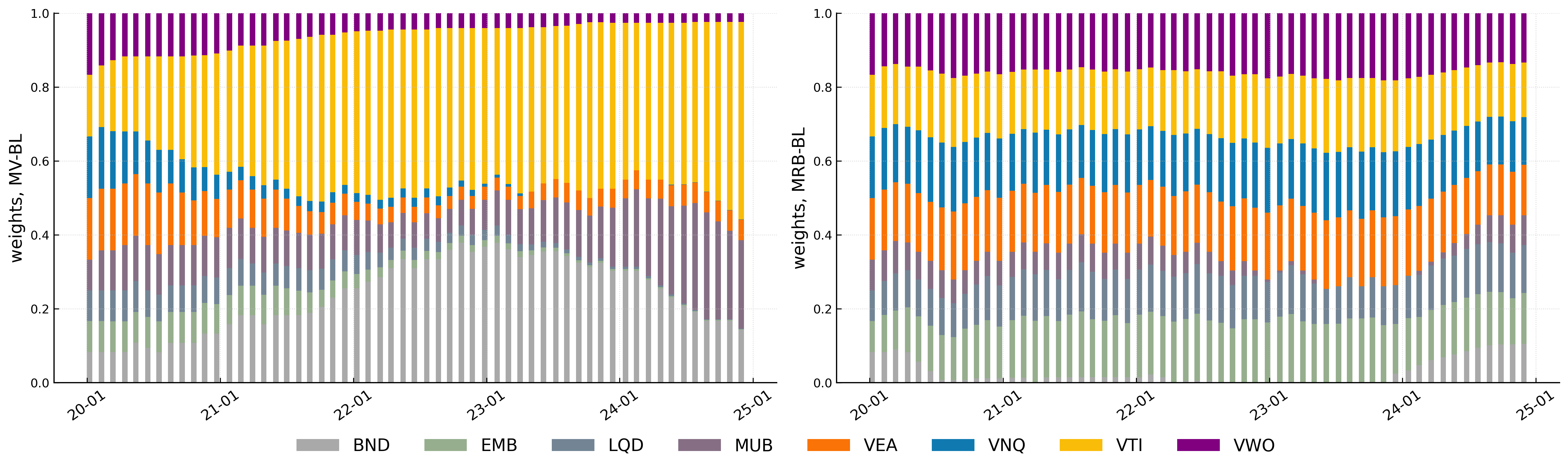}
	\caption{Portfolio weights for the MV-BL and MRB-BL strategies under a randomly varying risk attitude parameter $\gamma$. Top panel: unconstrained trading ($\delta = +\infty$); bottom panel: turnover-constrained case with $\delta = 0.05$.}
	\label{fig:random-gamma}
\end{figure}

We also conduct an experiment under a noisy risk-attitude profile, where $\gamma$ is sampled independently at each time step from the set $\{0.5, 0.7, 1, 1.5, 2\}$ and, correspondingly, $\gamma_R$ from the set $\{0.1, 0.2, 0.3, 0.4, 0.5, 0.6, 0.7, 0.8, 0.9, 1\}$. The results are shown in Figure~\ref{fig:random-gamma}.

In the absence of trading constraints, the portfolio weights fluctuate erratically and lack a coherent structure. Imposing a moderate turnover constraint ($\delta = 0.05$) significantly stabilizes the dynamics: the MRB-BL strategy behaves comparably to the case of a fixed risk attitude, indicating robustness to preference noise. In contrast, the MV-BL strategy remains highly sensitive to past realizations of $\gamma$, exhibiting no clear or stable pattern and instead reflecting persistent time spillover effects.

\subsubsection{Target portfolio}
We also perform a series of experiments imposing a practically relevant target portfolio constraint; see Section~\ref{sec:target-portfolio}. As in the previous section, we set the target risk attitude coefficient to $\gamma_{\mathrm{target}} = 0$, with all other parameters as in Section~\ref{sec:num-change-gamma}. The target constraint is activated over the final 14 rebalancing steps ($=2H$).

\begin{figure}[h]
	\centering
	\includegraphics[width=0.99\textwidth]{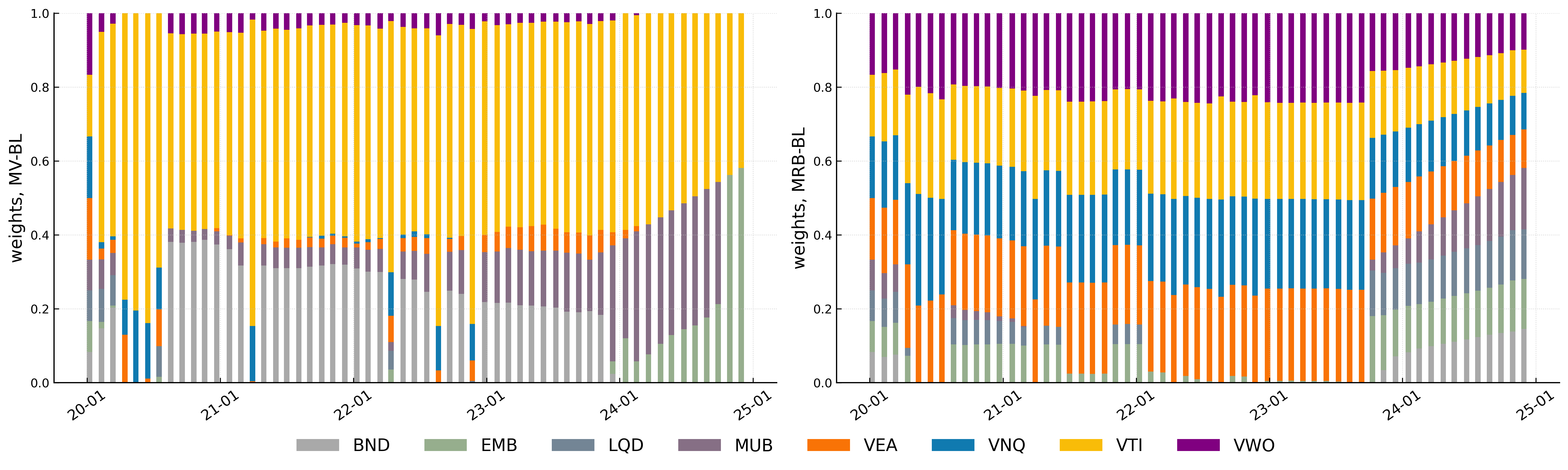}
		\includegraphics[width=0.99\textwidth]{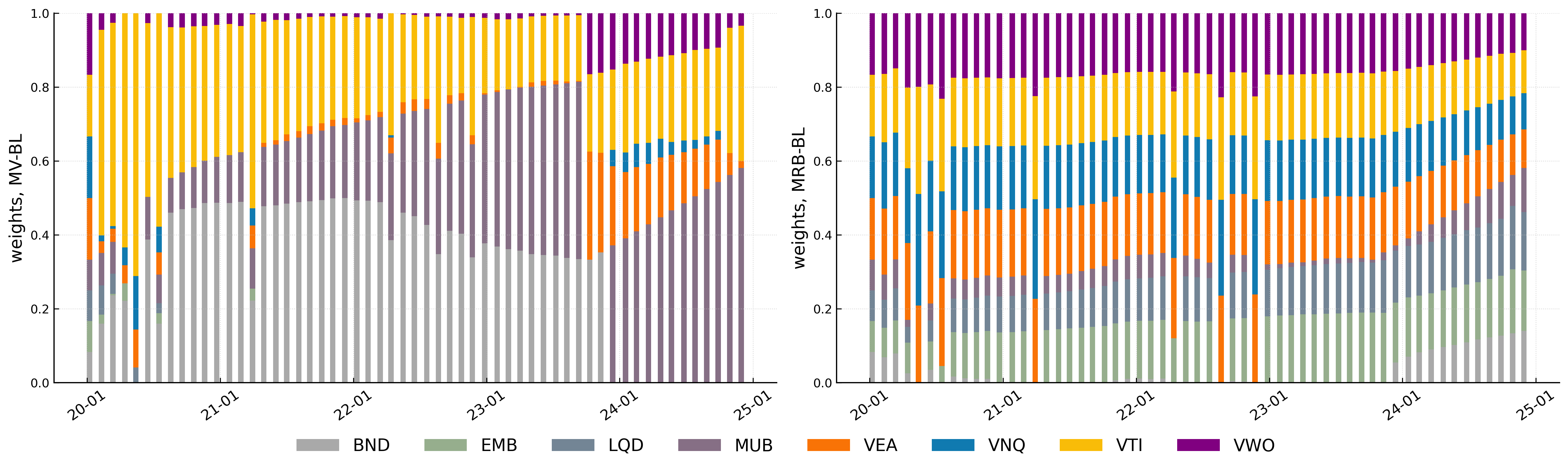}
	\caption{The weights for the MV-BL and MRB-BL strategies, under target portfolio and time varying risk attitude coefficient. Top panel: unconstrained trading ($\delta = +\infty$); bottom panel: turnover-constrained case with $\delta = 0.05$. }
	\label{fig:target-portf1}
\end{figure}

The results are presented in Figure~\ref{fig:target-portf1}, where $\gamma$ and $\gamma_R$ are either held constant (top panel) or vary linearly over time (bottom panel). For the static risk profile, the effect of the constraint is clearly visible: as maturity approaches, the portfolio weights adjust smoothly and monotonically to reach the prescribed target allocation.

In contrast, under a lifecycle profile for $\gamma$, the MV-BL strategy yields a portfolio that remains more diversified than in the unconstrained case, indicating a nontrivial interaction between time-varying risk attitudes and the target portfolio constraint. From a robo-advising perspective, this suggests that a prudent modeling approach is to combine the MV-BL framework with explicit target portfolio constraints and time-varying risk preferences, allowing for gradual risk de-risking while maintaining diversification and control near the investment horizon.

\section{Concluding remarks}\label{sec:conclusion}

Overall, the numerical analysis highlights substantial differences in how MV-Est-myopic, MV-Est-MPC, MV-BL and MRB-BL strategies respond to changes in risk attitudes, transaction cost controls, and MPC design parameters. Within the mean--variance framework, incorporating BL views and SMPC leads to more diversified and stable portfolios compared to classical estimated MV approaches, which exhibit sharp allocation switches and poor robustness with respect to the risk aversion parameter. At the same time, MV-BL remains sensitive to noisy or rapidly changing risk profiles, making explicit turnover constraints and target portfolio mechanisms essential for stabilizing portfolio dynamics in practical robo-advising implementations.

In contrast, the MRB-BL strategy produces markedly smoother and more stable allocations across a wide range of parameter choices, including $\gamma_R$, $\eta$, and $H$. While this structural stability may reduce turnover, it also limits responsiveness to market forecasts and renders soft transaction cost penalties largely ineffective, necessitating the use of hard constraints when trading activity must be controlled. Moreover, the limited sensitivity of MRB-BL to the MPC horizon suggests that the additional modeling and computational complexity of SMPC delivers relatively modest gains under risk-budgeting objectives.

Taken together, these findings suggest that a prudent robo-advising design should align the choice of optimization criterion, constraint structure, and preference modeling with the desired balance between adaptability and stability. In particular, MV-BL combined with explicit turnover or target portfolio constraints and time-varying risk attitudes offers a flexible and controllable framework, while MRB-based approaches may be better suited to settings where portfolio stability is prioritized and preference dynamics are slow or tightly regulated.

\section*{Acknowledgments}
The authors express their sincere gratitude to Tao Chen and Areski Cousin for their valuable input and support during the early phases of developing this manuscript and the underlying Python code. The authors acknowledge support from the National Science Foundation grant DMS-1907568.  IC also  acknowledges partial support from the National Science Foundation grant DMS-2407549.


\newcommand{\etalchar}[1]{$^{#1}$}

\end{document}